\documentclass{article}

\usepackage{arxiv}

\usepackage[utf8]{inputenc} 
\usepackage{CJKutf8} 
\usepackage[T1]{fontenc}    
\usepackage{hyperref}       
\usepackage{url}            
\usepackage{booktabs}       
\usepackage{amsfonts}       
\usepackage{nicefrac}       
\usepackage{microtype}      
\usepackage{graphicx}
\usepackage{doi}
\usepackage{amssymb}
\usepackage{array}

\newcommand{\tabref}[1]{Table~\ref{tab:#1}}
\newcommand{\figref}[1]{Figure~\ref{fig:#1}} 

\title{Task allocation interface design and personalization in gamified participatory sensing for tourism}

\author{ \href{https://orcid.org/0000-0002-3221-3612}{\includegraphics[scale=0.06]{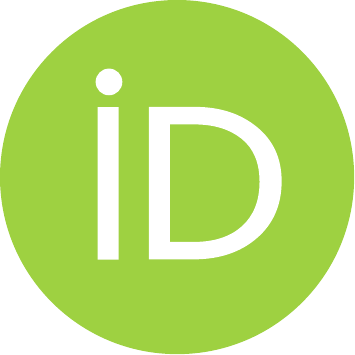}\hspace{1mm}Shogo Kawanaka} \\
	Nara Institute Science and Technology\\
	Takayama-cho 8916-5, Ikoma\\
	Nara, Japan, 630-0192 \\
	\texttt{kawanaka.shogo.kp1@is.naist.jp} \\
	\And
	Juliana Miehle \\
	Institute of Communications Engineering\\
	Ulm University\\
	Albert-Einstein-Allee 43, Ulm, Germany, 89081 \\
	\And
	Yuki Matsuda \\
	Nara Institute Science and Technology / Riken AIP\\
	Takayama-cho 8916-5, Ikoma\\
	Nara, Japan, 630-0192 \\
	\And
	Hirohiko Suwa \\
	Nara Institute Science and Technology / Riken AIP\\
	Takayama-cho 8916-5, Ikoma\\
	Nara, Japan, 630-0192 \\
	\And
	Keiichi Yasumoto \\
	Nara Institute Science and Technology / Riken AIP\\
	Takayama-cho 8916-5, Ikoma\\
	Nara, Japan, 630-0192 \\
	\And
	Wolfgang Minker \\
	Institute of Communications Engineering\\
	Ulm University\\
	Albert-Einstein-Allee 43, Ulm, Germany, 89081 \\
	
}

\renewcommand{\shorttitle}{Task allocation interface design and personalization in gamified participatory sensing}

\hypersetup{
pdftitle={Task allocation interface design and personalization in gamified participatory sensing for tourism},
pdfsubject={q-bio.NC, q-bio.QM},
pdfauthor={Shogo Kawanaka, Juliana Miehle, Yuki Matsuda, Hirohiko Suwa, Keiichi Yasumoto, Wolfgang Minker},
pdfkeywords={smart tourism, participatory sensing, interface design, communication style, personalization},
}

\begin{document}
\maketitle

\begin{abstract}
The collection of spatiotemporal tourism information is important in smart tourism and user-generated contents are perceived as reliable information. Participatory sensing is a useful method for collecting such data, and the active contribution of users is an important aspect for continuous and efficient data collection. This study has focused on the impact of task allocation interface design and individual personality on data collection efficiency and their contribution in gamified participatory sensing for tourism. We have designed two types of interfaces: a map-based with active selection and a chat-based with passive selection. Moreover, different levels of \textit{elaborateness} and \textit{indirectness} have been introduced into the chat-based interface. We have employed the Gamification User Types Hexad framework to identify the differences in the contributions and interface preferences of different user types. The results of our tourism experiment with 108 participants show that the map-based interface collects more data, while the chat-based interface collects data for spots with higher information demand. We also found that the contribution to sensing behavior and interface preference differed depending on the individual user type.
\end{abstract}


\section{Introduction}
The tourism industry has become a major industry accounting for 10.3\% of the world's total GDP in 2019, the demand is increasing every year, and services are needed to provide more comfortable tourism for the expansion of the industry in the future \footnote{Economic Impact | World Travel and Tourism Council (WTTC): \url{https://wttc.org/Research/Economic-Impact} (Accessed at 12, Feb., 2021)}.
Information on tourist destinations is very important when deciding on a destination~\cite{bib:iso-tour_smartcities_2020}.
With the development and spread of information technology, people can easily post and view their own tourism experiences, and the information generated by consumers through their actual experiences is accepted as a more effective and reliable source of information~\cite{bib:Sigala_Ashgate_2012}.
It is also changing tourism style, and a new of tourism, such as on-site tourism, which determines the next destination while sightseeing, is becoming popular~\cite{bib:dbj}.
Research on on-site tourism planning that takes into account the dynamically changing tourism context has been conducted, and the realization of such systems will provide a highly satisfying tourism experience~\cite{bib:hidaka_SmartCities_2020}.
However, the realization of such systems requires dynamic tourism information with high spatio-temporal resolution.
In order to collect such information, the participatory sensing approach~\cite{bib:burke_participatory_2006}, in which mobile devices owned by the general public such as smartphones are used as sensing devices, will be effective~\cite{bib:Gretzel_SmartTourism_EM_2015}.
While it has the property of collecting data with high spatio-temporal resolution at low cost, the problem is that the amount and quality of collected data depends on the contribution of the participating users.
As an incentive mechanism for improving user contributions and motivation to participate, gamification that incorporates game concepts and mechanisms into content other than games is attracting attention~\cite{bib:deterding_mindtrek_2011}.
Gamification has been applied in a variety of fields and has been shown to be effective so far~\cite{bib:ueyama_gamification_2014, bib:arakawa_gamification_2016}. According to literature on designing effective incentive mechanisms, points, rankings/leaderboards, and achievement are often introduced as gamification mechanics, and affect the user positively, particularly in participatory sensing~\cite{bib:jaimes_survey_ieee_2015, bib:seaborn_gam_survey_elsevior_2015, bib:Benedikt_IJHCS_2017}.
However, previous researches suggest that empirical studies are required in order to clarify the effects of gamification in each participatory sensing context or domain~\cite{bib:hamari_gamification_hicss_2014, bib:Benedikt_IJHCS_2017, bib:Sigala_Springer_2015}.
Additionally, the importance of individually personalizing the gamification design and the interface rather than the full package has also been revealed~\cite{bib:himari_homo_2013, bib:personality-gamification_jia_2016}.

In this paper, we investigate the effects of different task allocation interfaces and gamification user types on tourism information collection efficiency and tourism satisfaction in gamified participatory sensing for tourism.
We designed two types of task allocation interfaces, map-based and chat-based, and implemented them on our gamified participatory sensing platform application.
We also designed four different communication style sentences to investigate the appropriate way to ask the sensing tasks through the chat with agent character.
Furthermore, we employed the Gamification User Types Hexad framework to elucidate how an individual's personality influences their contribution to sensing and interface preference.
Then, we set these four main research questions:
\begin{description}
  \item[RQ1] How does the different task allocation interfaces affect the quantity and quality of dynamic tourism information collection?
  \item[RQ2] Do the different task allocation interfaces have an impact on tourism satisfaction of the tourists?
  \item[RQ3] Is there a relationship between tourism information collection efficiency and interface preference and gamification user type?
  \item[RQ4] What is the impact of different communication style sentences in a chat-based interface?
\end{description}
To elucidate these research questions, we conducted large scale experiment with 108 participants at actual sightseeing attractions in Nara, Japan.

As a result, we found that there was no difference in the effect of each interface on sightseeing satisfaction, but the characteristics of the collected data that, map-based allows for the collection of quantitative data and chat-based allows for the efficient collection of data needed by the system, were different. In addition, we found that different user types had different tendencies for their contribution to sensing and their interface preferences.
Moreover, 47~\% of the participants noticed a difference in the agent's communication style, showing that a large number of participants were aware of these subtle changes. Furthermore, the participants significantly preferred the elaborate communication style over the concise one. However, there is no preference for the \textit{indirectness} dimension. This shows that there is no general preference in the system's communication style and therefore the preference appears to be individual for every person.

The rest of this paper will describe the following.
First, we describe related work on the application of gamification in tourism and personalization in gamification. Then, we present the design and implementation of our designed two interfaces, and report the results of our experiments. Finally, we summarize this paper based on the results obtained from each experiment.

\section{Related Work}

\subsection{Gamification in Tourism}
Gamification is often introduced in the tourism domain in order to raise brand awareness, enhance tourist experiences, destination loyalty, consumer loyalty and engagements so far~\cite{bib:Sigala_Springer_2015, bib:Xu_TM_2017}.
However, according to Xu et al.~\cite{bib:Xu_TM_2017}, academic research on the application of gamification specifically in the tourism field is still scarce.
Ces\'{a}rio et al.~\cite{bib:teenage_vanessa_2020} designed two mobile apps, AR (Augmented Reality) game-based and story-based, to better understand teenage museum tourism behavior and to inform the suitable museum mobile app design. Their study also focused on user personalities and found that story-based strategies are suitable for a broader set of personalities.
There are a few researches investigating the impact of gamification on dynamic tourism information collection with various gamification mechanics. 
Although a comprehensive evaluation using various mechanics was conducted, it is not clear which gamification mechanism had an impact on which outcome~\cite{bib:lee_psy_behav_mdpi_2019}.
The prior work by Kawanaka et al. have previously investigated the impact of different task load and reward mechanisms on data collection efficiency and tourism behavior in a gamified participatory sensing application for tourism~\cite{bib:gamified_kawanaka_mdpi_2020}.
However, there is room to investigate the elements needed to personalize the design, such as the task allocation interfaces and user preferences.

\subsection{Interface and Interaction Design}

The notion that user interface design can be informed by other design practices has a rich tradition in HCI. During the first boom of computer games in the early 1980s, Malone wrote seminal papers deriving ``heuristics for designing enjoyable user interfaces'' from video games~\cite{bib:malone_motivating_1981}.
One of the studies on interface design in terms of data collection is the study of interfaces in online surveys, which was reported by Kim et al.~\cite{bib:kim_chatbot_2019}.
In this study, a web-based interface and a chatbot-based interface were used .They used formal style as basic style and casual style which is more friendly. The experimental results suggested, the quality of the responses was higher for chatbots than the web interface. Second, high quality answers were obtained in casual conversational style only in chatbots. From these results, it is suggested that differences in interface and conventional style in agent interaction affects the quality of data.
Another related study of communication styles is \textit{elaborateness} and \textit{indirectness} which was proposed by Pragst et al.~\cite{bib:Pragst_comstyle_2017}. \textit{Elaborateness} refers to the amount of additional information provided to the user and \textit{indirectness} describes how concretely the information that is to be conveyed is addressed by the speaker. Miehle et al.~\cite{bib:Miehle_usersatisfaction_2018} addressed the issues of how varying communication styles of a spoken user interface are perceived by users and whether there exist global preferences in the communication styles \textit{elaborateness} and \textit{indirectness}. It is shown that the system's communication style influences the user's satisfaction and the user's perception of the dialogue and that there is no general preference in the system's communication style. The authors conclude that spoken dialogue systems need to adapt their communication style to each user individually during every dialogue in order to achieve a high level of user satisfaction. In~\cite{bib:Miehle_estimation_2020}, a classification approach is presented addressing the estimation of the user's communication style in a spoken dialogue. It is shown that it is feasible to automatically detect the \textit{elaborateness} and \textit{indirectness} of the user during a dialogue. 


\subsection{Personalization in Gamification}
Several studies have explored the relations between gamification design and user motivational factors or personalities in human-computer interaction research area~\cite{bib:hamari_playertype_2014,bib:personality-gamification_jia_2016}.
Jia et al.~\cite{bib:personality-gamification_jia_2016} investigated the relationships among individuals’ personality traits and perceived preferences for various motivational affordances used in gamification. Their research showed correlations between each of the Big Five personality traits and the ten gamification factors.
Marczewski~\cite{bib:ninja_monkeys_2015} proposed the Hexad framework which has six gamification user types that differ in their motivational factors. The user types are personifications of people's intrinsic (e.g. self-realization) and extrinsic (e.g. rewards) motivations, as defined by the self-determination theory~\cite{bib:deci_sdt_2000}.
\tabref{hexad} shows the gamification user types defined by the Hexad framework as well as the motivations and characteristics of each user type~\cite{bib:ninja_monkeys_2015, bib:hexed-usertype_tondello_2019}.
Tondello et al.~\cite{bib:hexed-usertype_tondello_2016} proposed the 24-items survey response scale to score the user's preferences towards the six different motivations in the Hexad framework. This measure has the potential to accurately measure user preferences in gamification. There are four survey items related  to each user type, and all answers are rated on a 7-point Likert scale.
The score of each user type is obtained by adding the answers of each of the four questions, and the highest score is the user type.
Note that the Hexad user type is an archetypical categorization where the types represent users for whom certain motivations are stronger than other motivations.

\begin{table*}[t]
    \caption{Gamification user types defined in the Hexad framework}
    \label{tab:hexad}
    \centering
        \begin{tabular}{lll}
        \toprule
        User types & Motivation & Characteristics  \\ \hline
        \midrule
        Philanthropist & \begin{tabular}{l}Purpose\end{tabular} & \begin{tabular}{l}They are altruistic and willing to give without expecting a reward.\end{tabular} \\ \hline
        Socialiser & \begin{tabular}{l}Relatedness\end{tabular} & \begin{tabular}{l}They want to interact with others and create social connections.\end{tabular} \\ \hline
        Free Spirit & \begin{tabular}{l}Autonomy\end{tabular} & \begin{tabular}{l}They like to create and explore within a system. \\Freedom to express themselves and act without external control.\end{tabular} \\ \hline
        Achiever & \begin{tabular}{l}Competence\end{tabular} & \begin{tabular}{l}They seek to progress with a system by completing tasks, \\ or prove themselves by tackling different challenges.\end{tabular} \\ \hline
        Player & \begin{tabular}{l}Extrinsic \\ rewards\end{tabular} & \begin{tabular}{l}They will do whatever to earn a reward \\ within a system independently of the type of the activity.\end{tabular} \\ \hline
        Disruptor & \begin{tabular}{l}Triggering \\ of change\end{tabular} & \begin{tabular}{l}They tend to disrupt the system either \\ directly or through others to force negative \\ or positive changes.\end{tabular}\\ 
        \bottomrule
    \end{tabular}
\end{table*}

\section{Study Design}

In this study, we investigate the impacts of different task allocation interfaces on dynamic tourism information collection efficiency and tourism satisfaction in gamified participatory sensing.
In addition, we investigate whether the degree of the contribution to sensing and the interface design preferences are influenced by the individual personality.
We first describe the basic design, including the data to be collected and the assumed environment, in the next section.  After that, we describe the details of each interface and its implementation in the application.

\subsection{Basic design}
The data to be collected by participatory sensing for tourist and the assumed environment in this study are as follows.
The data is dynamic tourism information, specifically photos, comments and inertial sensor data built into smartphones at the sightseeing attractions.
With the development of human activity recognition research, it is possible to collect information such as tourist behaviors~\cite{bib:narimoto_wayfinding_2018} and congestion degree in the surrounding area~\cite{bib:Elhamshary_crowdmeter_percom_2018} from sensor data collected from smartphones.
Also, photos and comments are very useful information for the next tourist to understand the current situation of the sightseeing attraction. 
We use simple gamification mechanics, namely missions, point-based rewards and ranking.
Participants have our app installed on their own smartphones and posting photos and comments at a specific tourist attraction will appear on the app as a mission.
Each time they perform the mission while their sightseeing, the users will receive points as a reward within the app. Ranking is determined among the app users and displayed on the app, depending on the total amount of points.
Based on prior work by Kawanaka et al.~\cite{bib:gamified_kawanaka_mdpi_2020}, we use dynamic rewards that change according to the demand for information at each spot. This is designed under the assumption that campaign organizers (e.g. municipalities and tourism associations) can efficiently collect tourist information on the sightseeing spots they need. 
In this study, we designed two types of task (=mission) allocation interfaces based on these environments. 
One is a map-based interface in which the task is selected actively by the participant, and the other one is a chat-based interface in which the task is selected semi-passively based on the suggestion of the chat with agent character.

\subsection{Task Allocation Interfaces\label{interfaces}}

\begin{figure}[t]
    \centering
    \includegraphics[bb=0 0 403 219,width=0.9\columnwidth]{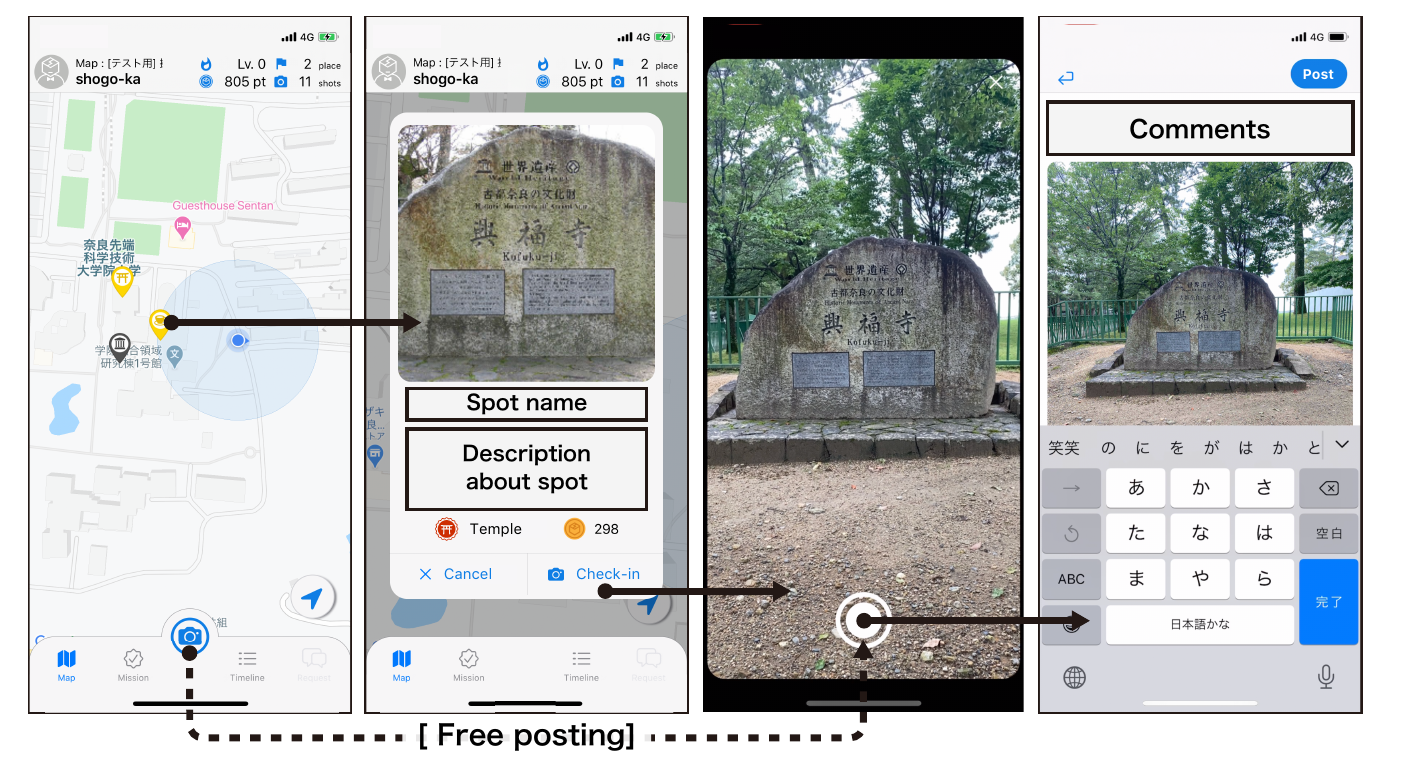}
    \caption{Map-based interface}
    \label{fig:screenshot_free}
\end{figure}

\noindent
\textbf{Map-based}\\
In this interface, all the spots to be sensed are displayed on the map and the user can accomplish the mission when actively checking the information of each spot one by one.
A screenshot of map-based interface is shown in \figref{screenshot_free}.
The pins displayed on the map are colored gold (high demand), silver (medium demand), and copper (low demand) according to the information demand level. The detailed information of each spot and the points are displayed by tapping on the pins.
The users can check-in by tapping the pin when they are a certain distance away from the target spot.
After that, the users can take a photo and post it with comment to complete the check-in process and receive a point.

\begin{figure}[t]
    \centering
    \includegraphics[bb=0 0 403 219, width=0.9\columnwidth]{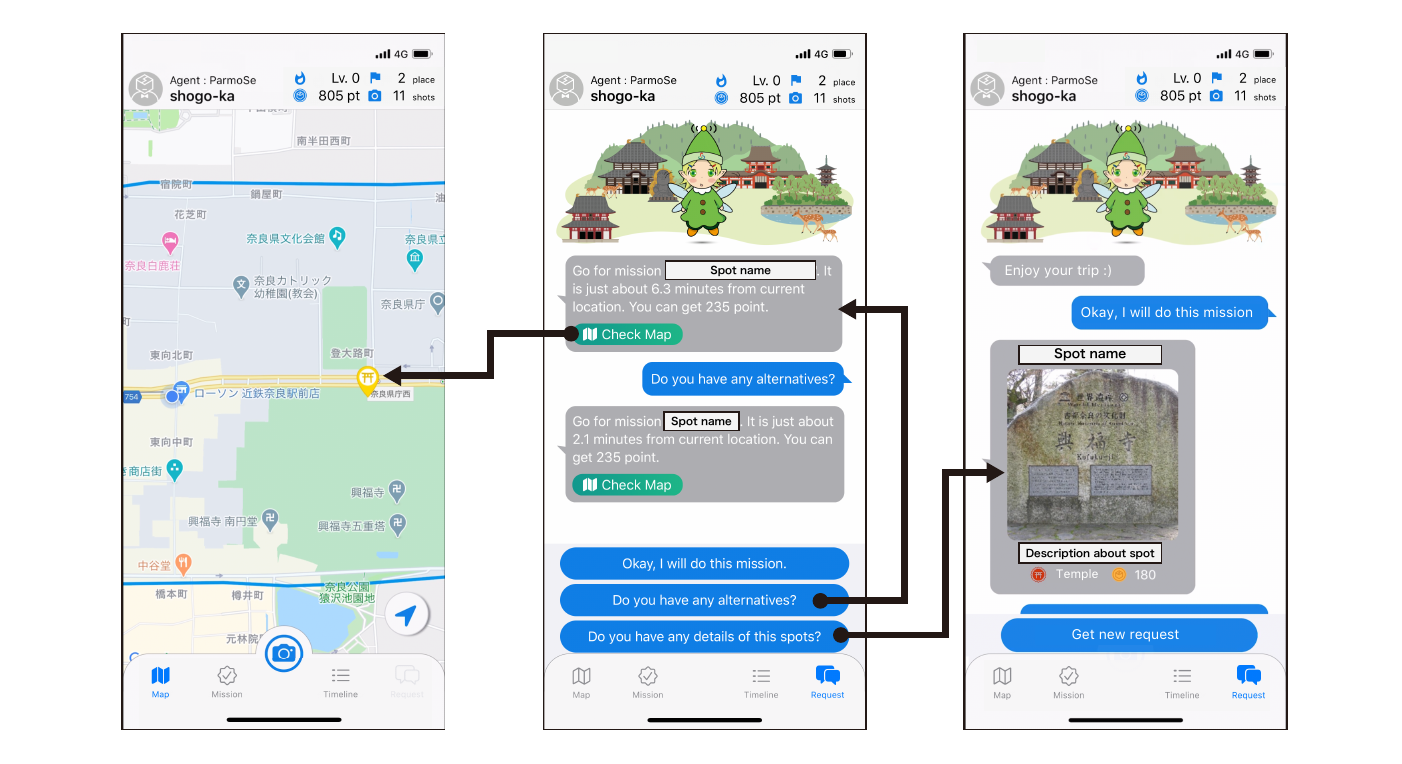}
    \caption{Chat-based interface}
    \label{fig:screenshot_agent}
\end{figure}

\noindent
\textbf{Chat-based} \\
In this interface, the main screen is a chat-based dialogue with the agent character and the user selects the mission during the interaction.
A screenshot of the chat-based interface is shown in \ref{fig:screenshot_agent}.
The agent asks the user to do a mission and the user accepts the specific mission in the dialogue and goes to the target spot to execute the mission.
The algorithm for determining the requested mission is as follows.
The user's current location is obtained and the linear distance between all the spots and the user is calculated. The ten closest spots are selected and sorted by points (the spot with the most points gets the highest priority).
At the beginning of the interaction, the agent selects the sightseeing spot with the highest priority (according to the algorithm) and asks the user to do this mission.
Then, the user has three choices (blue buttons at the bottom of the screen): 
\begin{description}
  \item[1] Okay, I will do this mission.
  \item[2] Do you have any alternatives?
  \item[3] Do you have any details of this spot?
\end{description}
In addition, by tapping the ``Check Map'' button on the chat screen, the specific locations can be seen on the map.
Through repeated dialogue, the users receives a mission that they want to do, go to the target spot and execute the mission.
The process of executing the mission is the same as for the map-based.

Additionally, we implemented \textit{Free posting} as a common function to both interfaces.
This allows participants to freely post any information they find of interest or want to share with other tourists while their sightseeing.
Free posting can be done by tapping the camera button at the bottom of the screen.

\subsection{Communication Style Design in Chat-based Interface \label{subsec:communication-style}}

As we mentioned in related research, when employing an interactive interface, the conventional style may affect the data quality and user's satisfaction.
In order to investigate the appropriate dialogue sentences and the effect on mission selection by the sentences, we formulated four different types of dialogue templates considering the communication styles \textit{elaborateness} and \textit{indirectness} in the chat-based interface.
The following templates were created (The Japanese notation is added, since the experiment was conducted with only Japanese participants):

\begin{figure}[t]
    \centering
    \includegraphics[bb=0 0 930 226, width=\columnwidth]{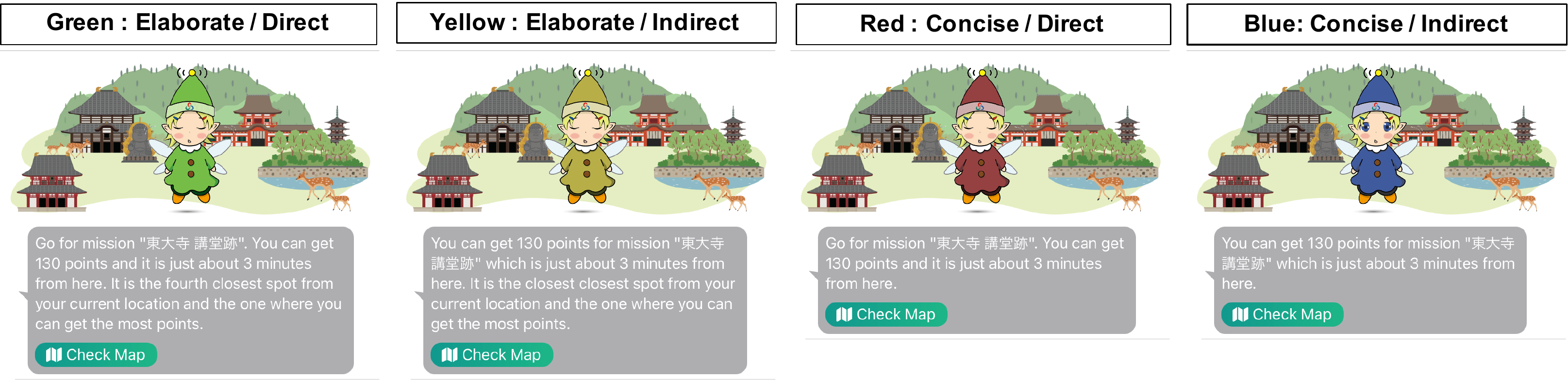}
    \caption{Correspondence of each communication style and agent character}
    \label{fig:communication_styles} 
\end{figure}

\begin{description}
    \item [Elaborate \& Direct(ED) ] : \vspace{0.5em}\\
    English : Go for mission <spot name>. You can get <100> points and it is just about <5> minutes from here. It is the <closest> spot from your current location and the one where you can get the <most> points. \vspace{0.3em}\\ 
    \begin{CJK}{UTF8}{min}
    Japanese : <スポット名> へ行ってください！ここから約 <5>で到着し、100 ポイントを獲得できます。現在地から <1>番目に近いスポットで、<1> 番目に多くポイントを獲得できるスポットです。
    \end{CJK}
    \item [Elaborate \& Indirect(EI) ] : \vspace{0.5em}\\
    English : You can get <100> points for mission <spot name> which is just about <5> minutes from here. It is the <closest> spot from your current location and the one where you can get the <most> points. \vspace{0.3em}\\
    \begin{CJK}{UTF8}{min}
    Japanese : ここから約 <5>分で到着する <スポット名> では <100> ポイントを獲得できます！ 現在地から <1> 番目に近いスポットで、<1> 番目に多くポイントを獲得できるスポットです。
    \end{CJK}
    \item [Concise \& Direct(CD) ] : \vspace{0.5em}\\
    English : Go for mission <spot name>. You can get <100> points and it is just about <5> minutes from here. \vspace{0.3em}\\
    \begin{CJK}{UTF8}{min}
    Japanese : <スポット名>へ行ってください！ ここから約 <5> 分で到着し、<100> ポイントを獲得できます。
    \end{CJK}
    \item [Concise \& Indirect(CI) ] : \vspace{0.5em}\\
    English : You can get <100> points for mission <spot name> which is just about <5> minutes from here. \vspace{0.3em}\\
    \begin{CJK}{UTF8}{min}
    Japanese : ここから約 <5> 分で到着する<スポット名> では、<100> ポイントを獲得できます！
    \end{CJK}
\end{description}

To make each interaction style more distinguishable, the colour of the agent's clothing changes with the communication style as can be seen in \figref{communication_styles}. 

\subsection{Application Implementation}
We implemented these task allocation interfaces into the participatory sensing platform application called Parmosense~\cite{bib:yukimat_parmosense_arXiv_2021}.
The timestamp, GPS, acceleration, gyroscopic, geomagnetism, and illuminance values of the smartphone are collected at a sampling rate of 10 Hz while this application is running (even in the background).
The data is sent to the server every 5 seconds.
Sensor data is also collected at the moment when the user takes a photo and is sent to the server along with the captured photograph, independently of the periodic sensor data.

\section{Experiment}

We conducted a tourism experiment to elucidate how the different interfaces and user types affect the efficiency of data collection, tourism satisfaction, and tourism behavior.
In this chapter, we first describe the recruitment procedure and the participants' demographic information. 
After that, we explain the experimental procedure and the contents of post-survey.

\subsection{Participants}

We recruited participants through research participant recruitment company.
Participants were limited to those who were Japanese, over 18 and under 80 years old, and living outside Nara Prefecture where the tourism experiment is conducted.
We asked to complete a questionnaire about the age, gender, previous tourism experience in the experimental area, and user types using the Hexad Gamification User Types Scale~\cite{bib:hexed-usertype_tondello_2019} during the application process. 
Finally, there were 157 applicants, and 110 participants were selected based on the results of the questionnaire on user types and tourism experience.
However, we describe the data and results from 108 participants because two of them could not collect the data normally.

\begin{figure}[htbp]
  \centering
    \begin{minipage}{\columnwidth}
        \centering
        \includegraphics[bb=0 0 889 328, width=0.9\columnwidth]{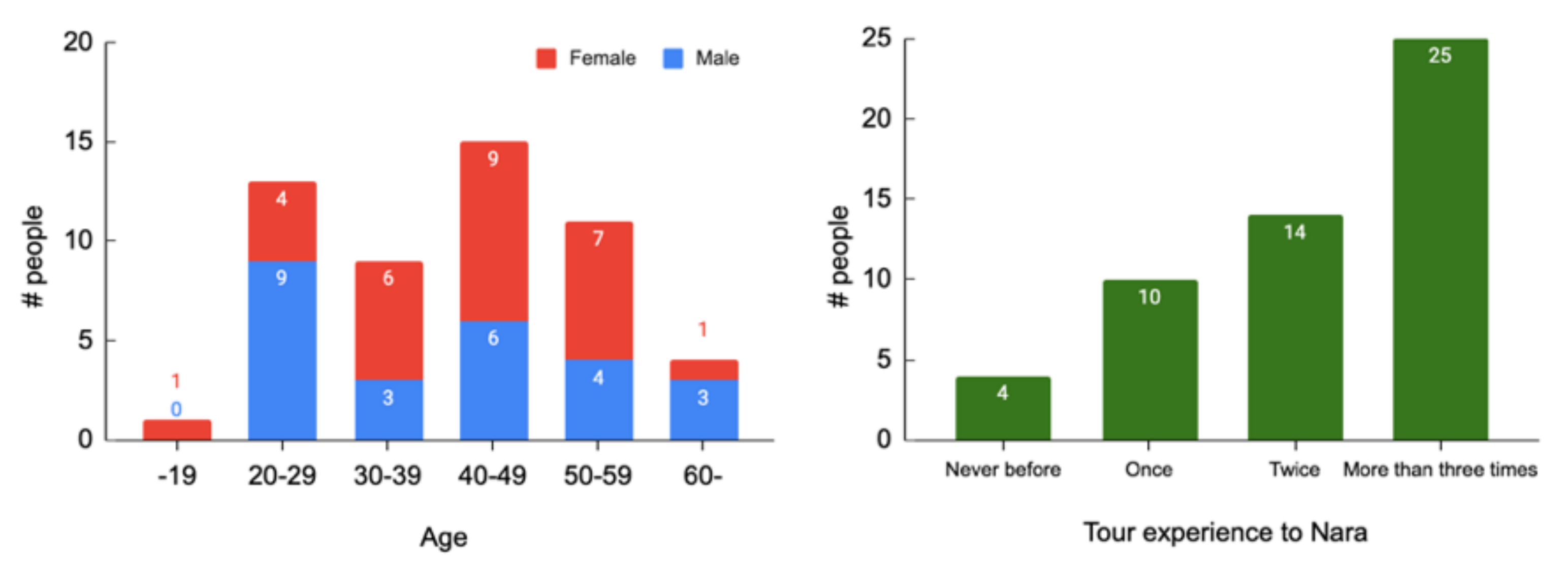} \\
        (a) GroupA : Map-based interface
      \end{minipage}
    \\ 
    \begin{minipage}{\columnwidth}
        \centering
        \includegraphics[bb=0 0 888 329, width=0.9\columnwidth]{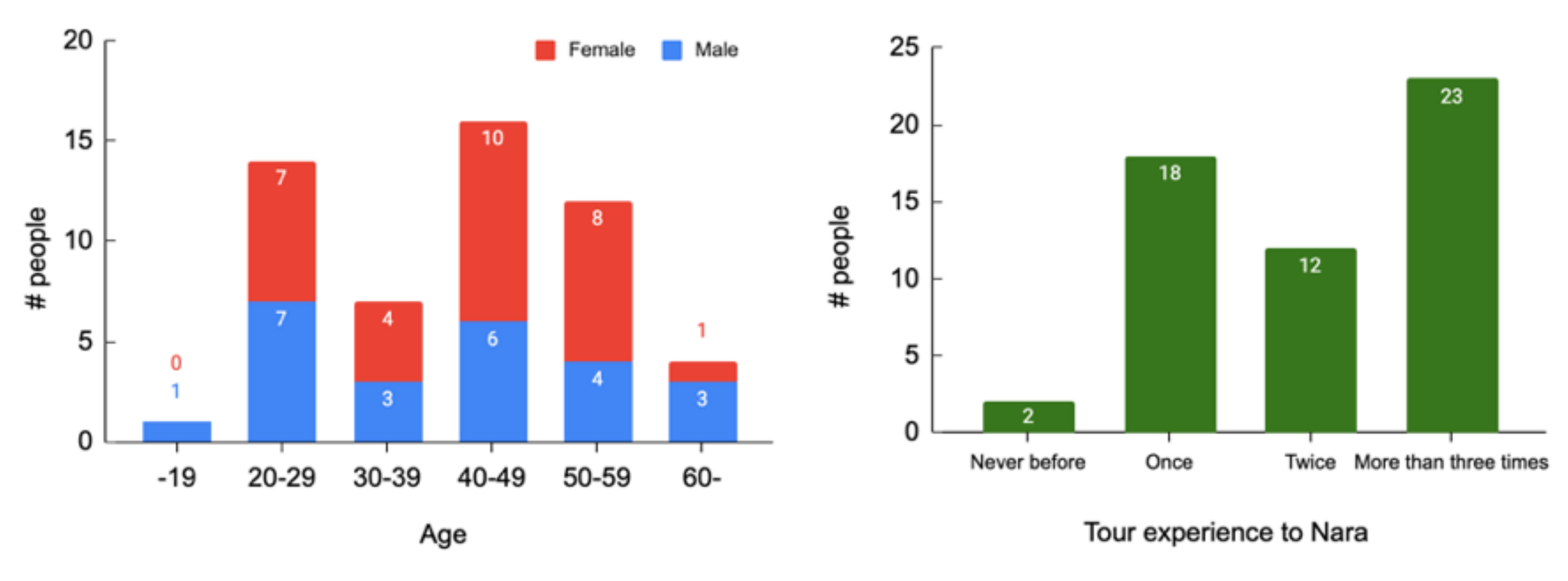} \\
        (b) Group B : Chat-based interface 
      \end{minipage}

    \caption{User attributes for each experimental group}
    \label{fig:lgex-user-attributes}
\end{figure}

There were 50 male and 58 female and aged between 19 and 71 years ($M = 41.0, SD = 13.9$).
With regard to the tourism experience, six people had never visited the area before, and 28 people had visited it once. 26 people had visited it twice and 49 people had visited it more than three times.
In terms of the Hexad gamification user types, 49 people as Free Spirit, 46 people were categorized as Philanthropist, 21 people as Player 17 people as Socialiser and 14 people as Achiever (There were some participants with multiple user types, the total number of user types exceeds 108.). Disrupter were not included in this experiment.
In this experiment, we divide the experimental group into two: one group using the map-based interface and the other group using the chat-based interface.
The distribution of males and females by age group and their tour experiences to Nara are shown in \figref{lgex-user-attributes}. 

\begin{table}[t]
    \centering
    \caption{Number of participants by experimental group for each date}
    \label{tab:participants-assignment}
    \begin{tabular}{c|wc{1.4cm}wc{1.3cm}wc{1.3cm}wc{1.3cm}wc{1.4cm}wc{1.3cm}}
    \toprule
    Date & Oct. 3rd & 26th & 28th & 31st & Nov. 1st & 4th. \\
    \midrule
    Group A & 4 & 10 & 11 & 11 & 9 & 8 \\
    Group B & 4 & 10 & 9 & 10 & 14 & 8 \\ \hline
    Total & 8 & 20 & 20 & 21 & 23 & 16 \\
    \bottomrule
    \end{tabular}
\end{table}

\subsection{Experimental Procedure}
The experiment was conducted in Nara, Japan in October 3rd, 26th, 28th, 31st, November 1st, and 4th, 2020\footnote{This experiment is carried out with the approval of ethics review committee of Nara Institute Science and Technology. In order to prevent the spread of COVID-19, participants were required to 1. wear a mask, 2. avoid long-term stays in crowded places, and 3. stop the experiment immediately if they were not feeling well.}. About 10 to 20 people were assigned to each of the above dates, taking into account the available dates of the participants. 
The number of participants for each dates were shown in \tabref{participants-assignment}.

Detailed explanation of the experiment was given to the participants through documents and any questions about experiment were accepted at any time.
The experimental application was installed on a participants' own smartphone in advance using TestFlight, a beta app delivery platform provided by Apple.
The use of the application was explained through documents and YouTube videos.
On the day of the experiment, the participants were gathered at the Kintestu-Nara station as start point of the experiment, and we explained the overview of the experiment, cautions on the experiment, and how to use the application one by one.
In addition, we answered individually to participants who had questions on the experiment and how to use the application.
The duration of the experiment was four hours, and the participants used the assigned interface at all times during the experiment.
In the main session of the experiment, we asked participants to do sightseeing alone and only on foot in the designated areas, while accomplishing missions and earning points. 
After the sightseeing experiment, participants answered to a post-survey about their tourism behavior and satisfaction, the usability of the application, and impressions throughout the experiment. The details of the post survey will be described in Section \ref{sec:lgex-post-survey}.
We paid each participant 8,000 yen ($\fallingdotseq$ 80 USD) as a basic participation fee including transportation fee to the venue. This is determined based on an hourly wage of 1,000 yen ($\fallingdotseq$ 10 USD), taking into account the 4 hours of experimental time, time for questionnaire response. In addition, since only people living outside Nara Prefecture were targeted in this experiment, an additional 3,000 ($\fallingdotseq$ 30 USD) yen was paid for round-trip transportation fee.

\subsection{Post-survey \label{sec:lgex-post-survey}}
We conducted a post-survey to subjectively evaluate tourism behavior and satisfaction and the usability of the application through the experiment.
The survey was divided into four categories: tourism satisfaction, interface preference and communication styles, application usability, and impressions throughout the experiment.

\noindent
\textbf{Tourism satisfaction}: The tourism satisfaction was assessed by the following questions on priorities between tourism and mission and the enjoyment of tourism.

\begin{description}
    \item[Q1] Which did you prioritize sightseeing or the mission?
    \item[Q2] Did the application make tourism more enjoyable for you?
\end{description}

For Q1, participants answered this question with a five-point Likert scale in which 1 $=$ prioritized sightseeing and 5 $=$ prioritized the mission, and and we asked the reason of their answers using an open question.
In Q2, they also answered this question with a five-point Likert scale in which 1 $=$ Not at all fun and 5 $=$ Very fun, and we asked the reason of their answers using an open question.

\noindent
\textbf{Interface preference and communication styles}: In this experiment, each participant used only one interface, so we asked them about their preference for the interface app.
For the communication style, we asked the participants who used chat-based interface application if they noticed any differences in sentences and which communication style they preferred, as follows.

\begin{description}
    \item[Q3] Do you like the map-based/chat-based style user interface?
    \item[Q4] Did you notice any changes to the sentence or appearance in your interactions with agents?
    \item[Q5] Which agent did you think was the best? Please pay attention to the sentence and answer it.
\end{description}

For Q3, participants answered this question with a five-point Likert scale in which 1 $=$Don't like it at all and 5 $=$Like it very much, and and we asked the reason of their answers using an open question.
Q4 was answered with binary option, yes or no, and Q5 was answered 1-4 options with sample screenings shots pasted for each communication style; 1: Elaborate \& Direct, 2: Elaborate \& Indirect, 3: Concise \& Direct, 4: Concise \& Indirect.

\noindent
\textbf{Application usability}: The application usability of each interface was evaluated with System Usability Scale~\cite{bib:sus}, which allows us to easily get the usability score (min:0, max:100) of a system with ten questionnaire items.

\noindent
\textbf{Impressions throughout the experiment}: Finally, we asked them for their impressions throughout the tourism experiment with open questions.

\section{Results}
Through the whole experiment, approximately 1.53 GB of sensor data, 3148 photos and comments, and 108 post-survey answers were collected.
In order to investigate the impact of different task allocation interfaces on tourism information collection efficiency, tourism behavior and tourism satisfaction in participatory sensing for tourists, a quantitative evaluation using the collected log data and a qualitative evaluation using the questionnaire results were conducted.

\subsection{Quantitative Evaluation using Collected Data}

\begin{table}[t]
    \centering
    \begin{tabular}{c|wc{1.8cm}wc{1.8cm}wc{1.8cm}|wc{2cm}|wc{2cm}}
    \toprule
    & \multicolumn{3}{c|}{Demand Level} & Free & \\ \cline{2-4}
    & Low & Middle & High & Posting & Total \\
    \midrule
    Total Posts & 562 & 470 & 480 & 290 & 1802 \\ \hline
    Average & 10.60 & 8.87 & 9.06 & 5.47 & 34.00 \\
    Median & 10 & 8 & 8 & 2 & 32 \\
    SD & 4.21 & 5.03 & 5.60 & 7.51 & 15.37 \\
    \bottomrule
    \end{tabular}
    \caption{Summary of the collected posts in map-based interface}
    \label{tab:summary-mission-map}
\end{table}

\begin{table}[t]
    \centering
        \begin{tabular}{c|wc{1.8cm}wc{1.8cm}wc{1.8cm}|wc{2cm}|wc{2cm}}
        \toprule
        & \multicolumn{3}{c|}{Demand Level} & Free & \\ \cline{2-4}
        & Low & Middle & High & Posting & Total \\
        \midrule
        Total Posts & 166 & 258 & 532 & 390 & 1346 \\ \hline
        Average & 3.02 & 4.69 & 9.67 & 7.09 & 24.47 \\
        Median & 2 & 4 & 9 & 2 & 22 \\
        SD & 3.00 & 3.24 & 5.52 & 10.91 & 14.77 \\
        \bottomrule
        \end{tabular}
    \caption{Summary of the collected posts in chat-based interface}
    \label{tab:summary-mission-agent}
\end{table}

\begin{figure}[t]
    \centering
    \includegraphics[bb= 0 0 381 186, width=0.6\columnwidth]{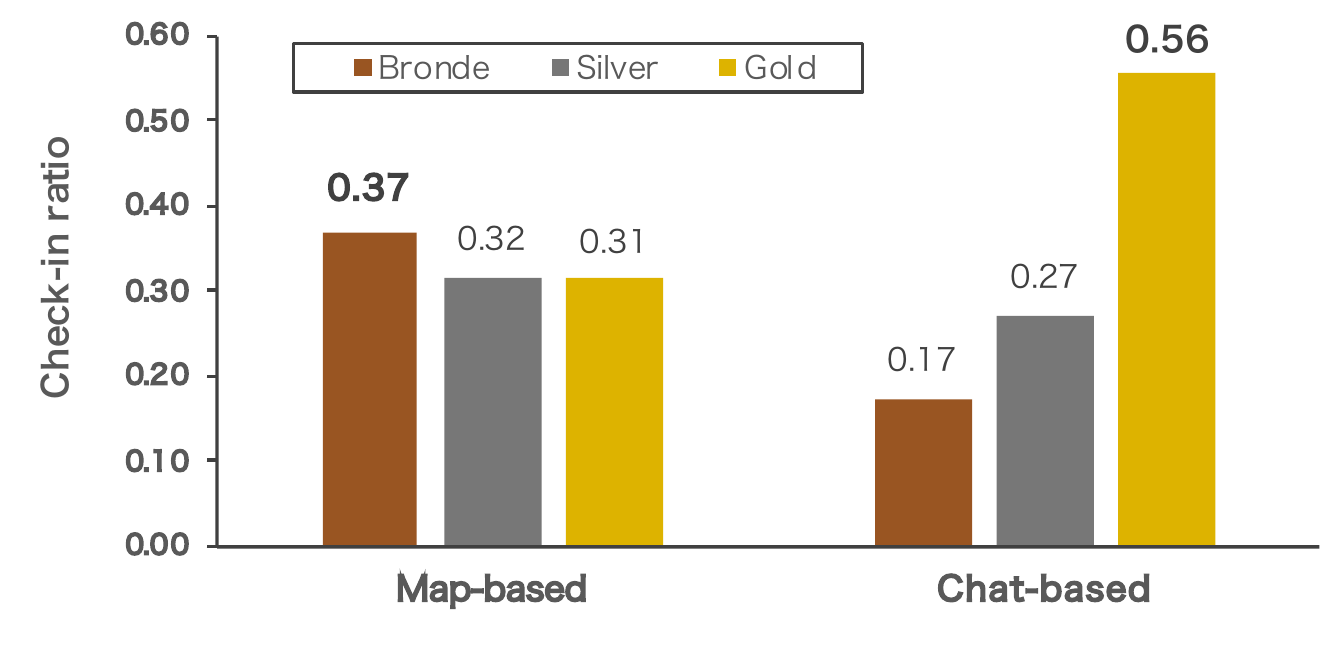}
    \caption{Check-in ratio according to the information demand level in large scale experiment}
    \label{fig:lgex-checkin-ratio}
\end{figure}

The summary of the collected posts in the map-based interface and chat-based interface are shown in \tabref{summary-mission-map} and \tabref{summary-mission-agent}, respectively.
The 1512 posts (average 28.5 posts per person) are submitted as check-in missions, the 290 posts (average 5.5 posts per person) are submitted as free posting and 1802 posts are obtained in total from map-based interface.
On the other hand, the 956 posts (average 17.4 posts per person) are submitted as check-in missions, the 390 posts (average 7.1 posts per person) are submitted as free posting and 1346 posts are obtained in total from chat-based interface.
The absolute number of posts obtained by the check-in mission was about 1.55 times greater on average for the map-based interface.
However, the absolute number of submissions obtained by free posting was about 1.42 times greater on average for the chat-based interface.
When both were combined, the posts was about 1.39 times greater on average for the map-based interface.

A statistical significance test was performed to determine if there is a significant difference for differences in the number of posts obtained for each interface. 
First, we performed a Shapiro-Wilk test to check the normality of the number of posts obtained in the check-in mission and free posting at each interface.
The results show that the total number of accomplished missions for the chat-based interface follows normality ($p=0.22$), but the rest does not follow.
We therefore ran a Mann-Whitney U Test to compare the number of missions completed by each participant and the number of free postings by each participant.
As a result, we found significant differences only between the posts obtained for the check-in missions in each interface.
We found that the number of posts from the check-in missions was significantly greater in the map-based interface.
On the other hand, the average number of posts obtained from free posting was grater for the chat-based interface, but the significant difference was not found.

\begin{figure}[t]
    \centering
        \begin{tabular}{c}
        \begin{minipage}{0.45\columnwidth}
            \centering
            \includegraphics[bb=0 0 383 343, width=\columnwidth]{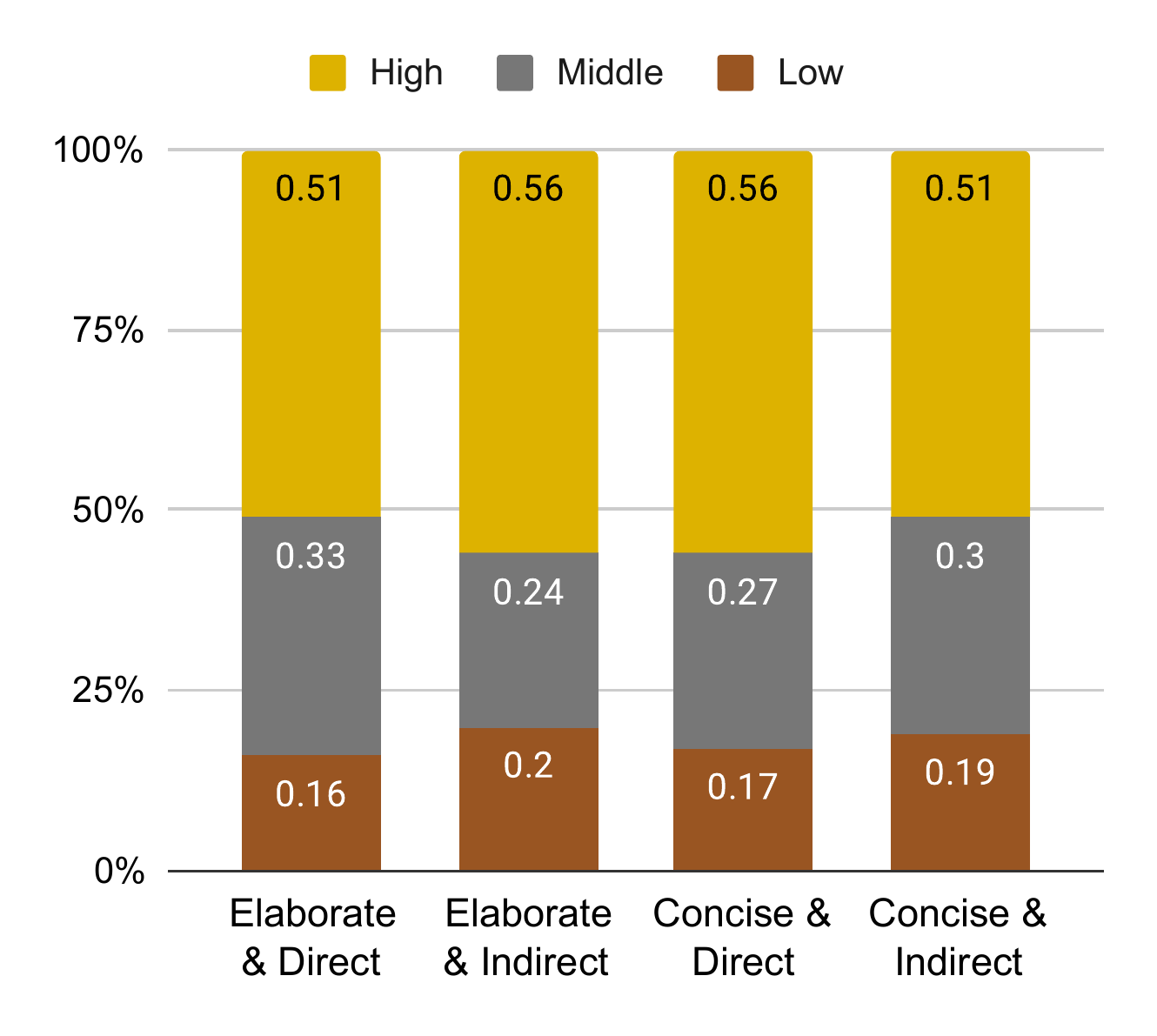}
            \hspace{1.6cm} {\small(a) Four communication styles }
        \end{minipage}
        
        \begin{minipage}{0.45\columnwidth}
            \centering
            \includegraphics[bb=0 0 383 343, width=0.98\columnwidth]{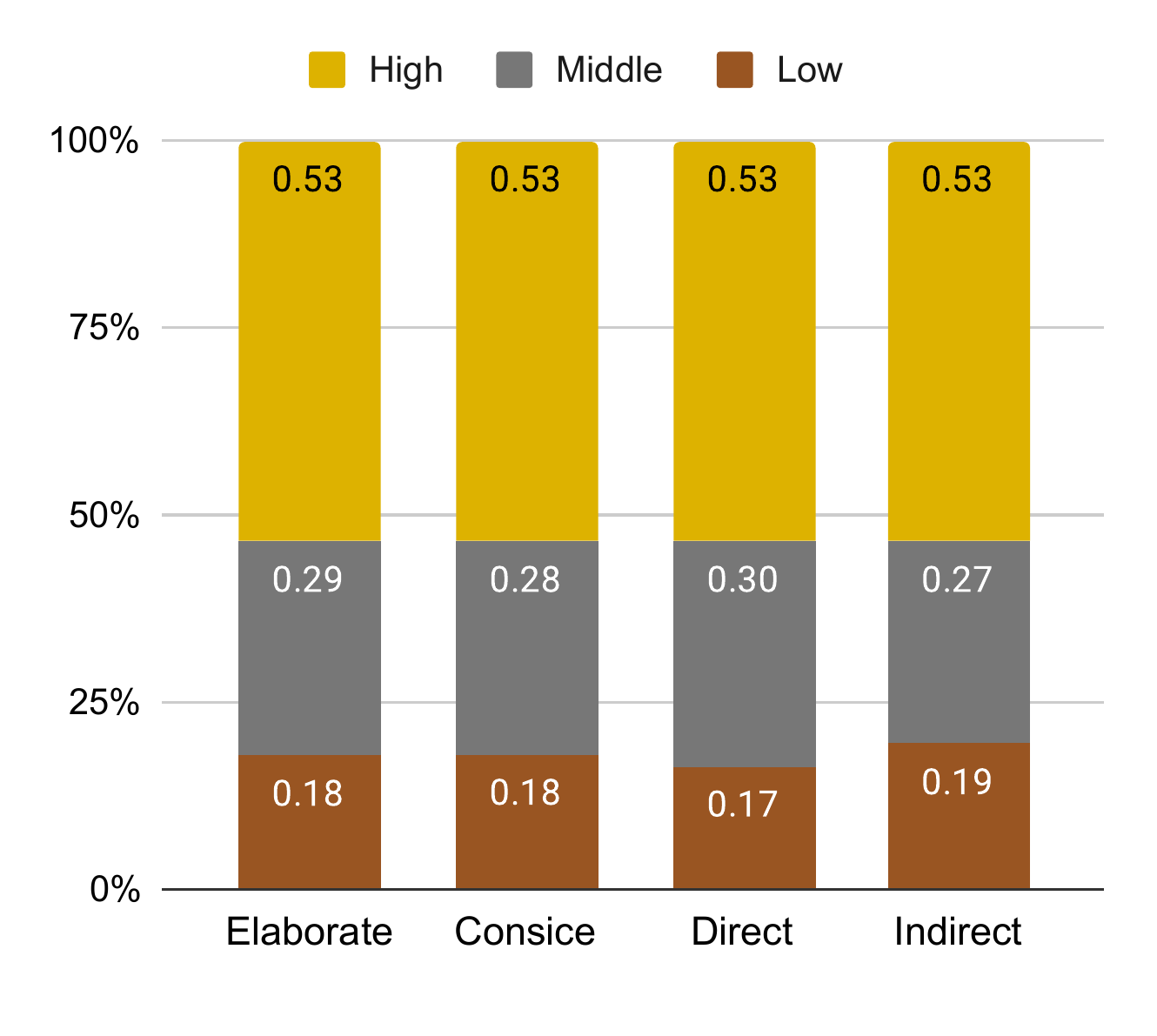}
            {\small(b) Elaborateness and Indirectness}
        \end{minipage}
    \end{tabular}
    \caption{Check-in rate by communication style in chat-based interface}
    \label{fig:lgex-chieckin-ratio-com}
\end{figure}

Next, the quality of the data was evaluated by the check-in ratio according to the information demand of each sightseeing attraction. The check-in ratio according to information demand level in each interface is shown in \figref{lgex-checkin-ratio}.
In the map-based interface, the rate was the highest for the low-demand spots, which are colored with bronze and assigned to more famous spots.
On the other hand, the participants were more likely to complete in the high-demand spots colored with gold.

Finally, we will elucidate whether the efficiency of data collection differs depending on the communication style in the chat-based interface.
\figref{lgex-chieckin-ratio-com} shows the check-in ratio for each communication style; the results are shown for each of the four communication styles in (a), and the results are tabulated for each dimension of Elaborateness and Indirectness in (b).
In \figref{lgex-chieckin-ratio-com} (a), the ratio of high demand missions in Elaborate \& Indirect and Direct \& Concise is slightly higher, but similar results are obtained in almost all styles.
In addition, similar results were also obtained for all items when we see the results from the two dimensions in (b).


\begin{figure}[t]
    \centering
        \begin{tabular}{c}
        \begin{minipage}{0.55\columnwidth}
            \centering
            \includegraphics[bb=0 0 361 220, width=\columnwidth]{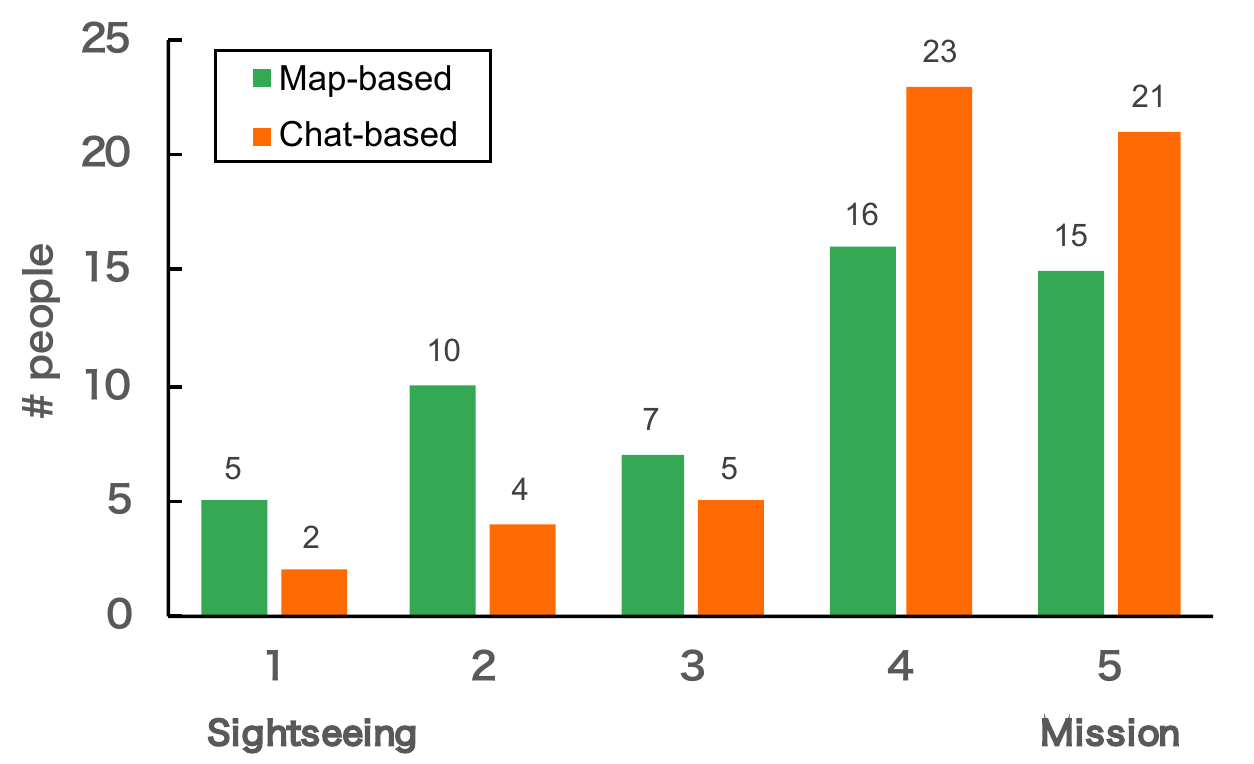}
            \hspace{1.6cm} {\small(a) The distribution of answers}
        \end{minipage}
        
        \begin{minipage}{0.35\columnwidth}
            \centering
            \includegraphics[bb=0 0 219 217, width=0.93 \columnwidth]{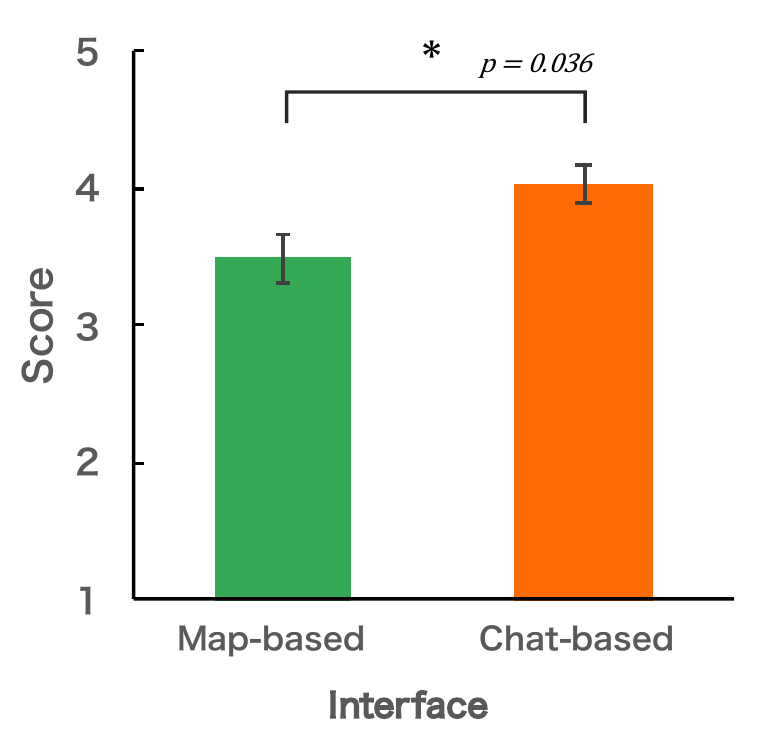}
            \hspace{1.6cm} {\small(b) Average score}
        \end{minipage}
    \end{tabular}
    \caption{Summary of the answers to Q1 on the priority of mission and sightseeing}
    \label{fig:lgex-priority}
\end{figure}

\begin{figure}[t]
    \centering
        \begin{tabular}{c}
        \begin{minipage}{0.55\columnwidth}
            \centering
            \includegraphics[bb=0 0 361 219, width=\columnwidth]{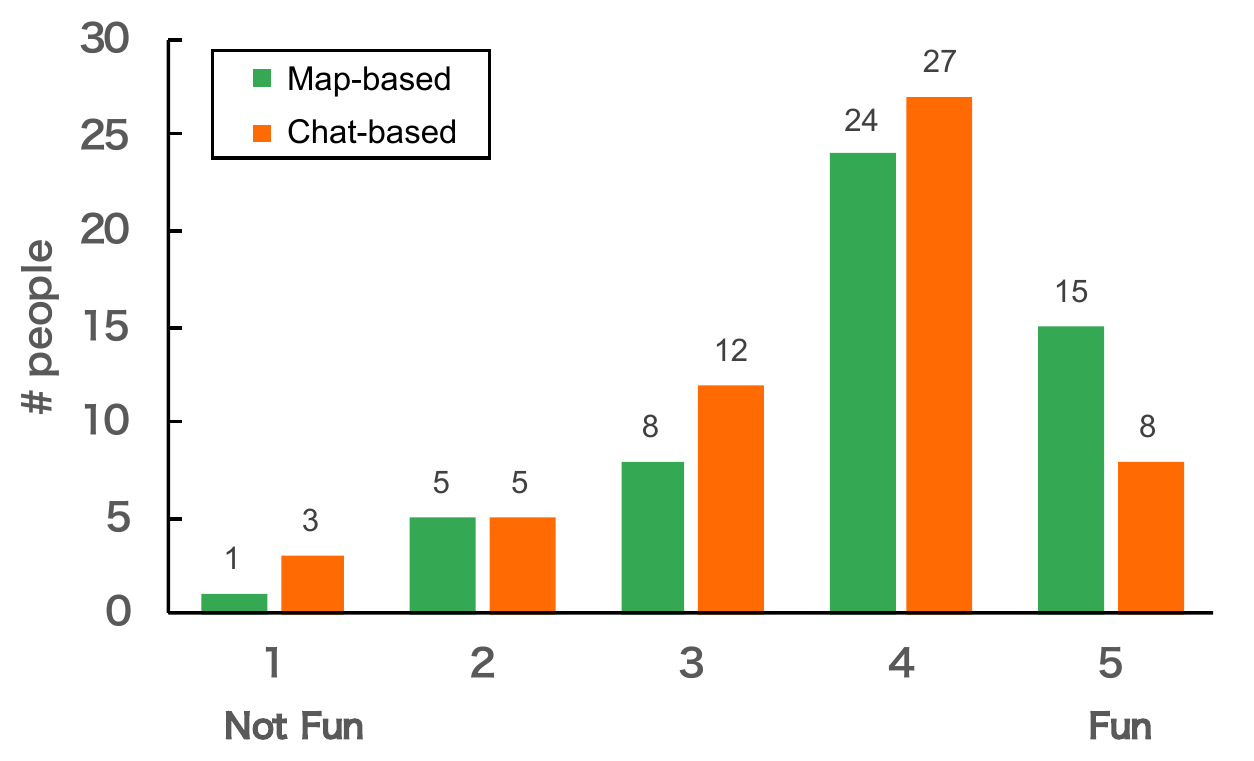}
            \hspace{1.6cm} {\small(a) The distribution of answers}
        \end{minipage}
        \begin{minipage}{0.35\columnwidth}
            \centering
            \includegraphics[bb=0 0 219 219, width=0.93\columnwidth]{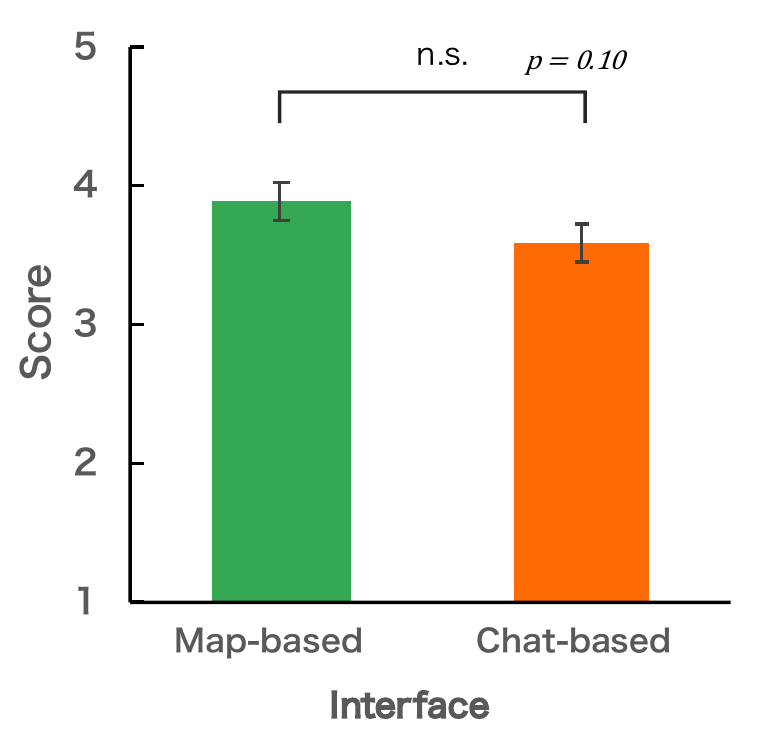}
            \hspace{1.6cm} {\small(b) Average score}
        \end{minipage}
    \end{tabular}
    \caption{Summary of the answers to Q2 on the sightseeing enjoyment}
    \label{fig:lgex-tour-enjoyment}
\end{figure}

\subsection{Quantitative Evaluation using Post-survey}

\textbf{Tourism satisfaction}:
The summary of the answers to Q1 on priorities between tourism and mission is shown in \figref{lgex-priority}; (a) shows the distributions of answers for each interface and (b) shows the average score of the answers.
The average and median scores for the map-based interface were 3.44 and 4.00 ($S.D. = 1.37$), and for the chat-based interface were 4.04 and 4.00 ($S.D. = 1.05$).
This result shows that missions are prioritized over tourism in both interfaces.
It was also found that there is a tendency to prioritize missions over map-based interface in the chat-based interface.
To determine if this difference is statistically significant, we performed a significance test.
The normality of the answers to Q1 for each interface was confirmed by the Shapiro-Wilk test, and the results both did not follow a normal distribution. Therefore, we used the Mann–Whitney U test, and we found a significant difference between them ($p < 0.05$).
We found that tourists are significantly more likely to prioritize missions with the chat-based interface than with the map-based interface.

Next, we summarize the results obtained by free description of the reasons for the responses.
In the map-based interface, the most common reason given by the participants who answered 5 or 4, i.e., who more prioritized the mission, was the ``Gameplay'', with 17 participants answering.
For example, the participant P71 (Female, 46) responded, \textit{``I wanted to go to the place with the highest points as much as possible because it became fun like a game.''} and the participant P66 (Male, 48) answered, \textit{``I was focusing on sightseeing until the middle of the tour, but since I'm the type of person who wants to compete when I'm given a ranking, so I became mission-oriented from the middle.''}
The second most common answer was ``For sightseeing reference'', with 6 respondents.
On the other hand, 15 participants answered, ``For sightseeing reference'', which was the most common reason for prioritizing missions in the chat-based interface.
For instance, the participant P29 (Female, 22) responded, \textit{``I thought there would be many places that I could only learn about through the app,''} and Participant P49 (female, 21) responded, \textit{``I didn't know this area well, so I followed the mission to go sightseeing.''}.
The second most common answer was ``Gameplay'', with 10 respondents.
In addition, as a characteristic answer of the chat-based, 8 people answered ``Sense of duty'', and as an example, the participant P47 (Female, 45) answered \textit{``Because the agent character asked me the mission.''}.

These responses indicate that in the map-based interface, gamification elements such as points and ranking were the factors that made people prioritize the mission more. In the chat-based interface, the passivity of asking a spot and the algorithm of prioritizing minor spots with high information demand were found to be factors.

The answers to Q2 on sightseeing enjoyment are summarised in \figref{lgex-tour-enjoyment}; (a) shows the distribution of answers for each interface and (b) shows the average score of the answers.
The average and median scores for the map-based interface were 3.87 and 4.00 ($S.D. = 0.99$), and for the chat-based interface were 3.58 and 4.00 ($S.D. = 1.03$).
This result shows that both interfaces make tourism more enjoyable.
It was also found that the map-based interface tend to make tourism more fun than the chat-based interface.
Similarly, we conducted a significance test to determine if there is a significant difference.
We used the the Mann–Whitney U test, since the results of the Shapiro-Wilk test did not follow a normal distribution.
The results of test showed that there is no significant difference between the enjoyment of sightseeing with the chat-based interface and with the map-based interface ($p = 0.10$).
That is, the enjoyment of tourism does not differ significantly between the different interfaces.


Next, we summarize the reasons for the answers regrading enjoyment.
The most common answer for the participants who answered 5 or 4, i.e., who responded that they enjoyed sightseeing more than usual, was ``Chance encounter'', with 17 and 22 participants in the map-based and chat-based, respectively.
For example, Group A participant P70 (Male, 69) responded, \textit{``I went to places I would not have normally gone to, but there were spots nearby. This leads to awareness.''} and Group B participant P58 (Male, 22) answered \textit{``I was able to visit minor places that I would not have chosen on my own.''}.
The reason for the larger number of respondents in the chat-based interface, it is assumed that lesser-known spots are preferentially requested by agent characters.
The second most mentioned factor was ``Gameplay'', with 9 and 6 respondents, respectively.
The examples are follows, Group A participant P76 (Male, 24) mentioned, \textit{``I thought I would sightseeing would be neglected, but I felt a sense of accomplishment by visualizing the trip with points and recorded the places I visited.''} and Group B participant P102 (Male, 39) answered, \textit{``Normally, it takes time to decide a tourist spot, but I felt that I was able to go around a lot by making it a mission by this application. ''}.


\begin{figure}[h]
    \centering
        \begin{tabular}{c}
        \begin{minipage}{0.55\columnwidth}
            \centering
            \includegraphics[bb=0 0 361 219, width=\columnwidth]{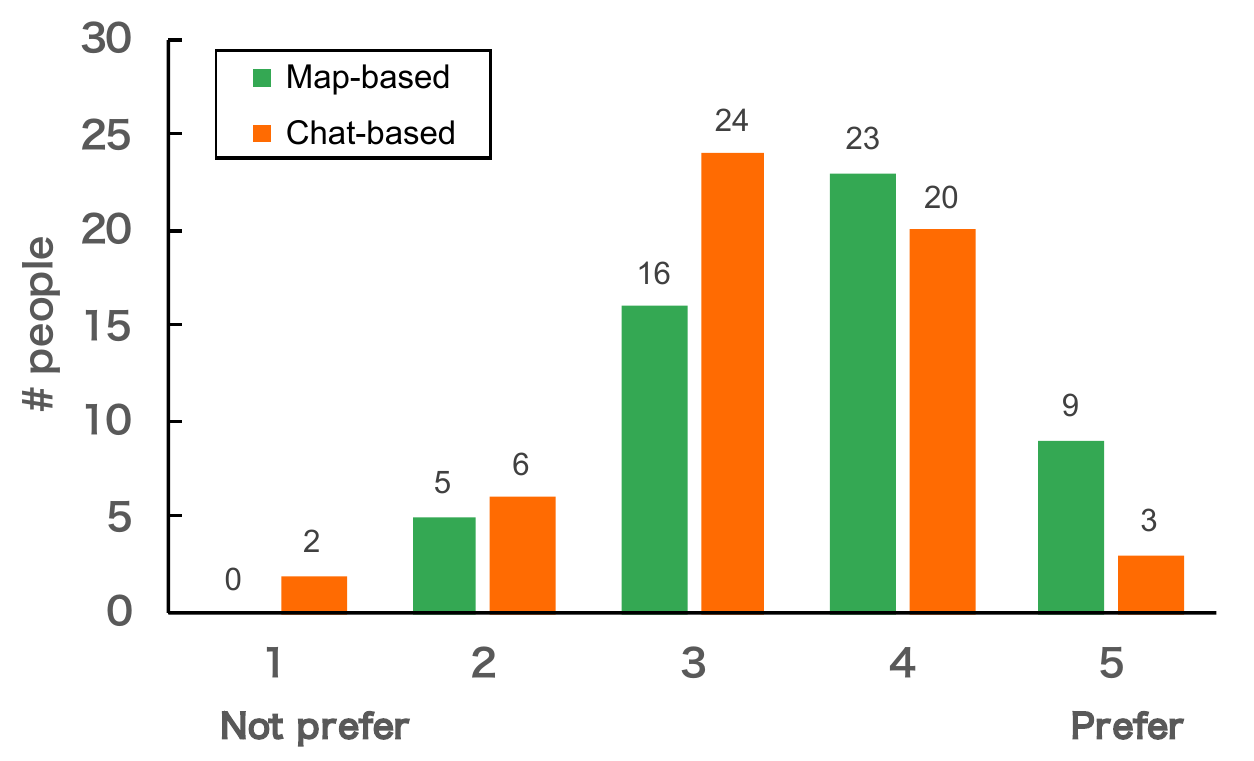}
            \hspace{1.6cm} {\small(a) The distribution of answers}
        \end{minipage}
        \begin{minipage}{0.35\columnwidth}
            \centering
            \includegraphics[bb=0 0 220 219, width=0.93\columnwidth]{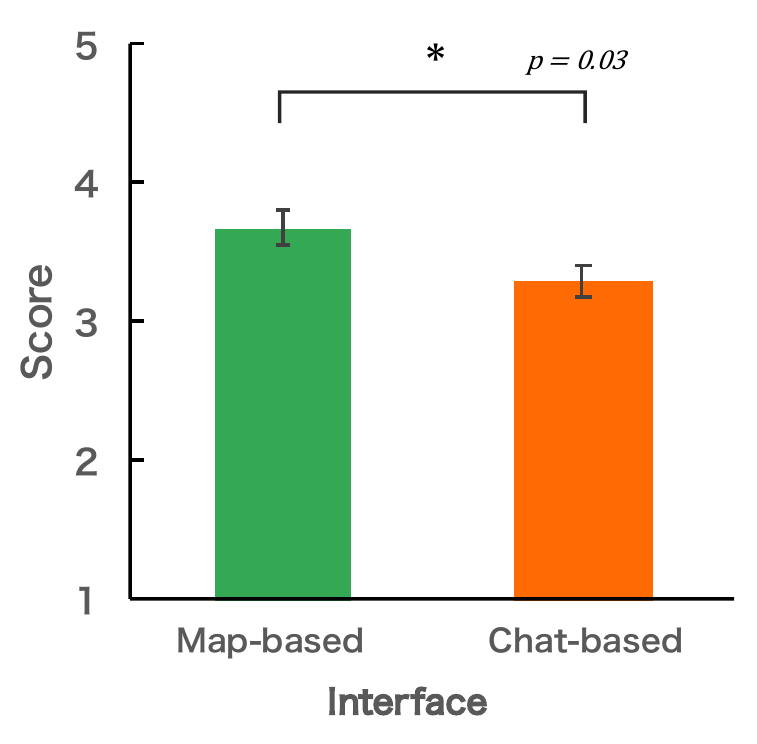}
            \hspace{1.6cm} {\small(b) Average score}
        \end{minipage}
    \end{tabular}
    \caption{Summary of the answers to Q3 on the interface preference}
    \label{fig:lgex-preference}
\end{figure}

\begin{figure}[h]
    \centering
    \includegraphics[bb= 0 0 385 311, width=0.47\columnwidth]{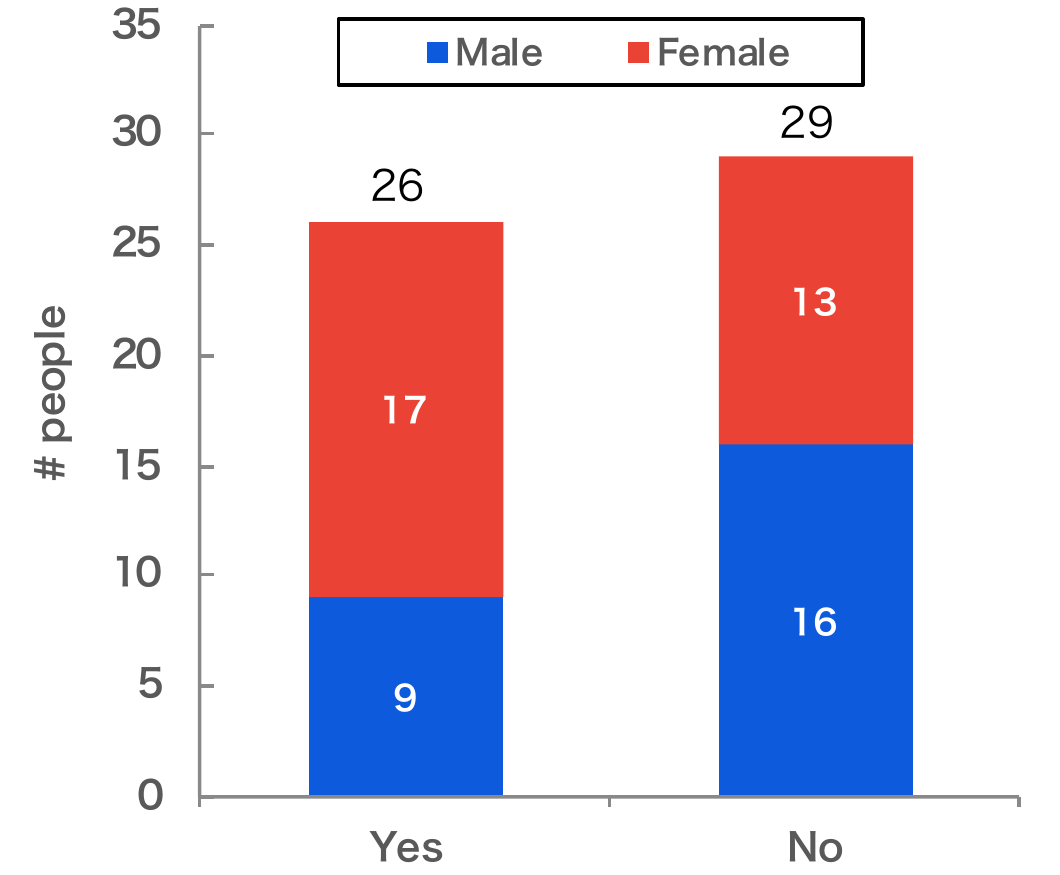}
    \caption{The answers to Q4 on whether the participants noticed the differences of sentence and appearance of agent characters}
    \label{fig:lgex-comstyles-notice}
\end{figure}

\noindent
\textbf{Interface and communication style preferences}:
The summary of the answers to Q3 on interface preference is shown in \figref{lgex-preference}. The mean $M$ and the median $Md$ for the map-based interface are $M(m)=3.69$ and $Md(m)=4.00$, the results for the chat-based interface are $M(c)=3.29$ and $Md(c)=3.00$. There is a statistically significant difference between $M(m)$ and $M(c)$ ($p=0.03$ using the Mann-Whitney-U-Test), showing that the participants prefer the map-based interface over the chat-based interface.

26 participants (47~\%) noticed a difference in the agent's interaction, while 29 participants (53~\%) did not notice any difference. The gender distribution is shown in \figref{lgex-comstyles-notice}. This result indicates that the female participants were more sensitive to the changes in the agent's communication style. However, there is no significant difference (using a Chi-Squared Test).


\begin{figure}[t]
    \centering
        \begin{tabular}{c}
        \begin{minipage}{0.5\columnwidth}
            \centering
            \includegraphics[bb=0 0 361 252, width=\columnwidth]{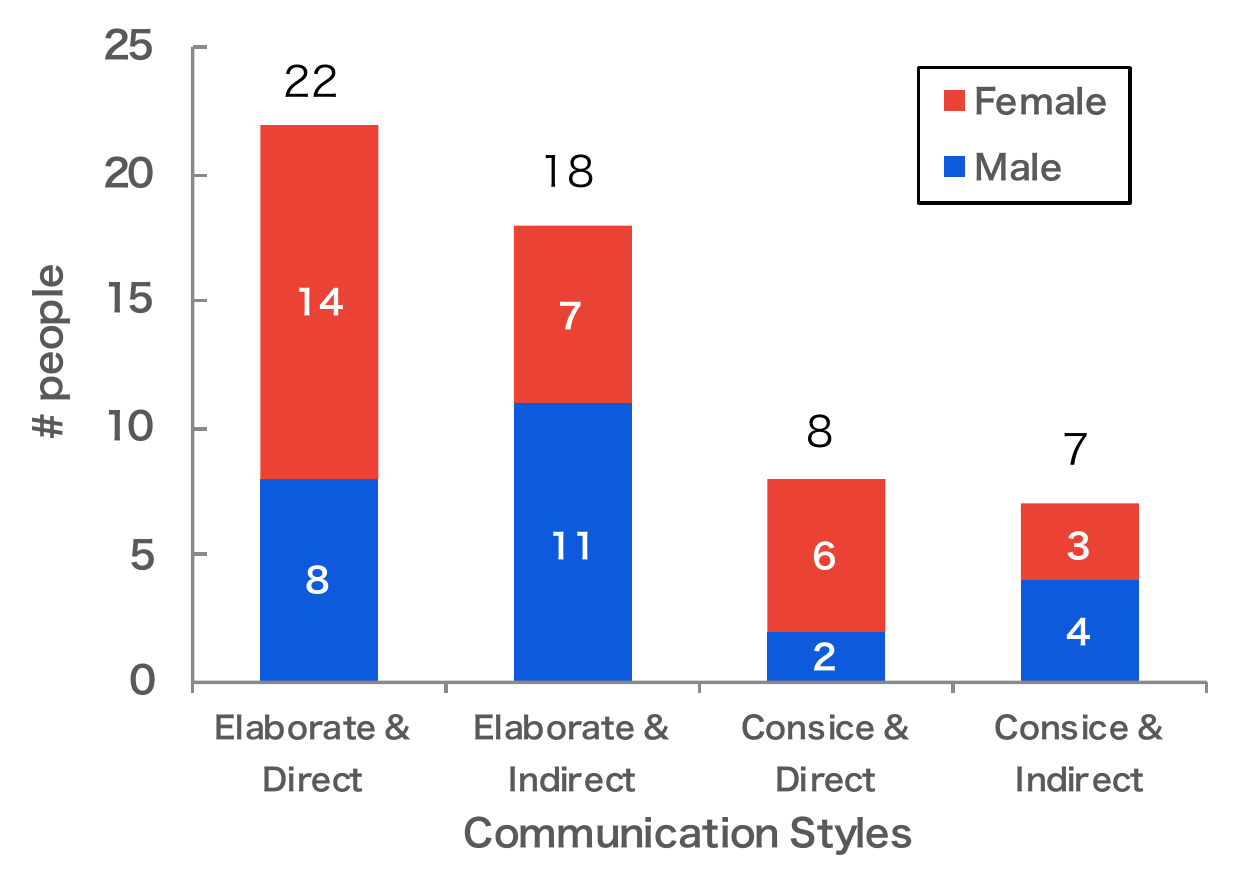}
        \end{minipage}
        \begin{minipage}{0.4\columnwidth}
            \centering
            \includegraphics[bb=0 0 316 252, width=\columnwidth]{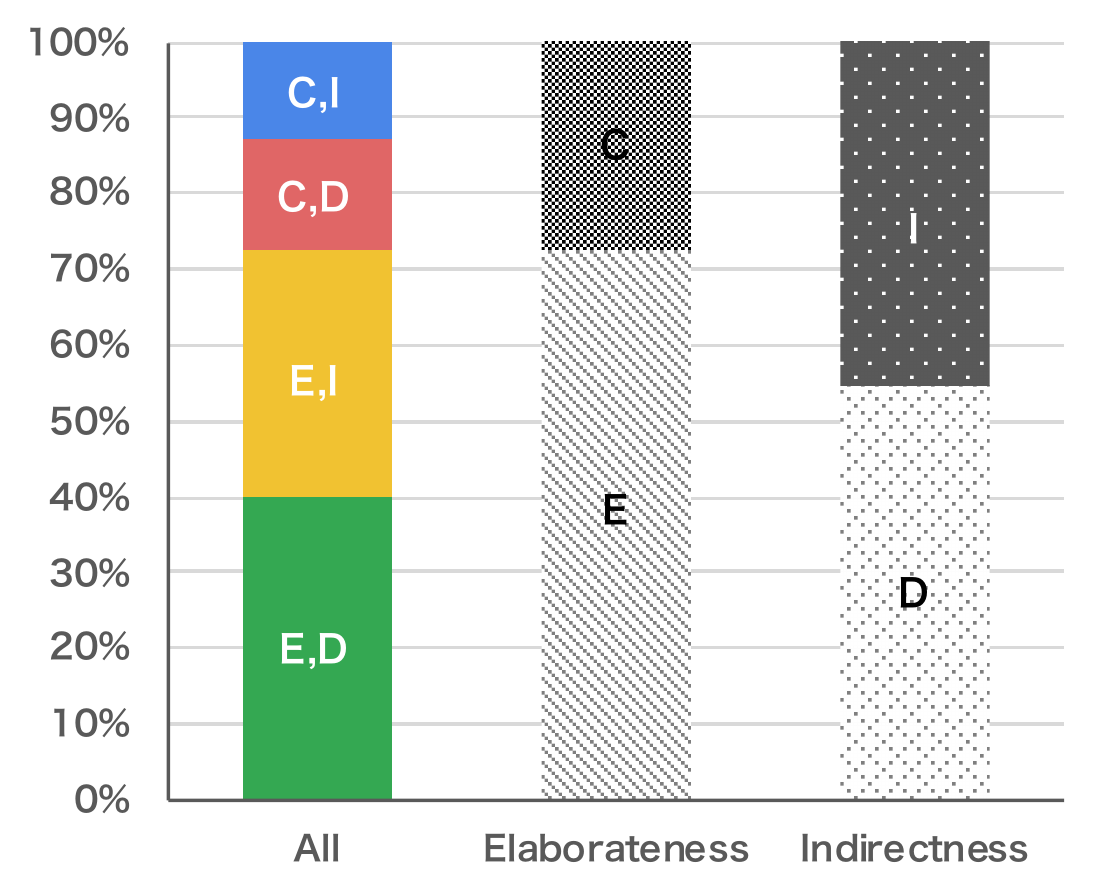}
        \end{minipage}
    \end{tabular}
    \caption{The result of Q5, showing the communication style preferences of all participants ($N=55$).}
    \label{fig:lgex-comstyles-all}
\end{figure}

\begin{figure}[t]
    \centering
        \begin{tabular}{c}
        \begin{minipage}{0.5\columnwidth}
            \centering
            \includegraphics[bb=0 0 361 252, width=\columnwidth]{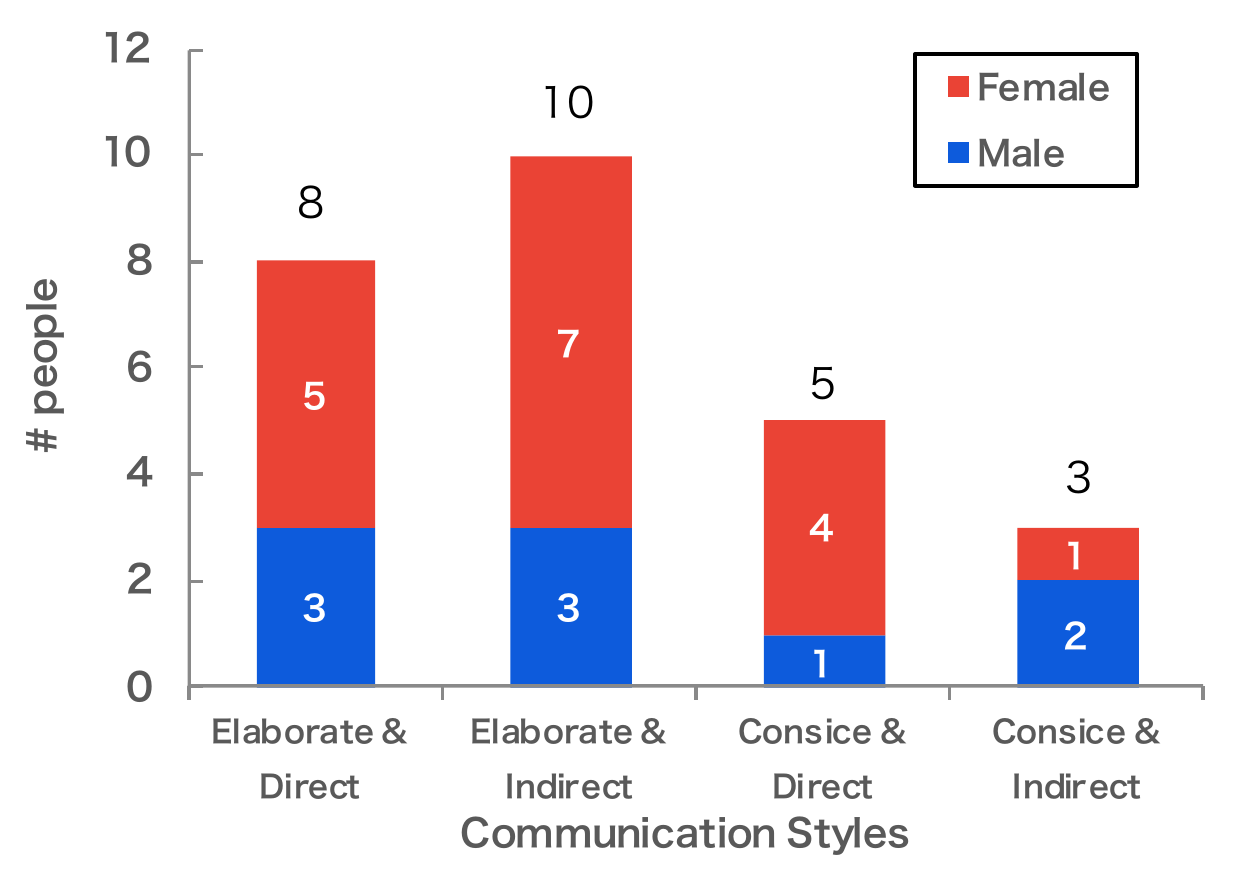}
        \end{minipage}
        \begin{minipage}{0.4\columnwidth}
            \centering
            \includegraphics[bb=0 0 316 252, width=\columnwidth]{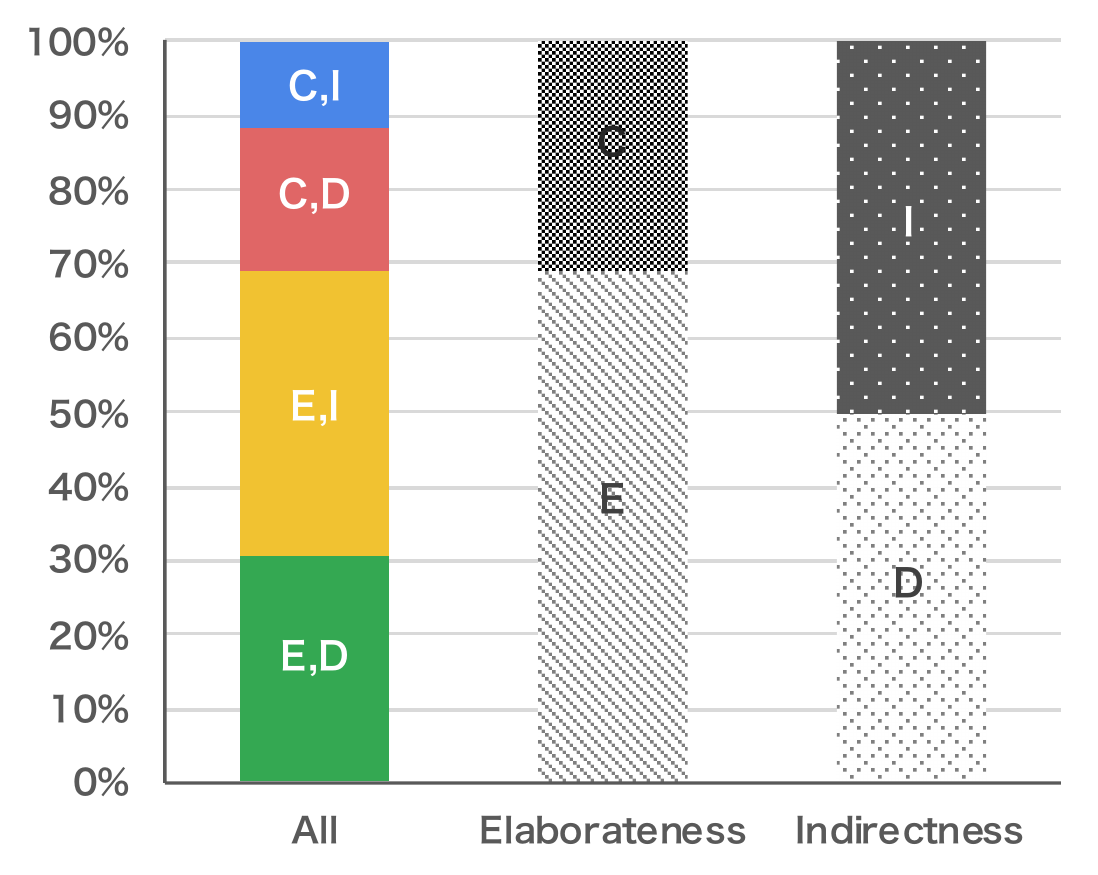}
        \end{minipage}
    \end{tabular}
    \caption{The result of Q5, showing the communication style preferences of all participants who noticed a difference in the agent's interaction ($N=26$).}
    \label{fig:lgex-comstyles-yes}
\end{figure}


\figref{lgex-comstyles-all} shows the communication style preferences of all participants ($N=55$) and \figref{lgex-comstyles-yes} shows the communication style preferences of all participants who noticed a difference in the agent's interaction ($N=26$). It can be seen that there is a clear preference in the \textit{elaborateness} dimension, both when looking at all participants and when looking at only those participants who noticed a difference in the agent's interaction. A Chi-Squared Test shows that the difference between elaborate and concise is statistically significant for both groups ($p<0,05$). Hence, the participants significantly preferred the elaborate communication style over the concise one. This shows that, in general, more detailed information is preferred for the task at hand. However, it has to be noted that this does not apply to all participants. Some of them clearly stated that they find a short and simple text easier to read on the smartphone while walking. Moreover, there is no preference for the \textit{indirectness} dimension. This is in line with the results presented in~\cite{bib:Miehle_usersatisfaction_2018}, that there is no general preference in the system's communication style and therefore the preference appears to be individual for every person.

\begin{figure}[t]
    \centering
        \begin{tabular}{c}
        \begin{minipage}{0.58\columnwidth}
            \centering
            \includegraphics[bb=0 0 362 219, width=\columnwidth]{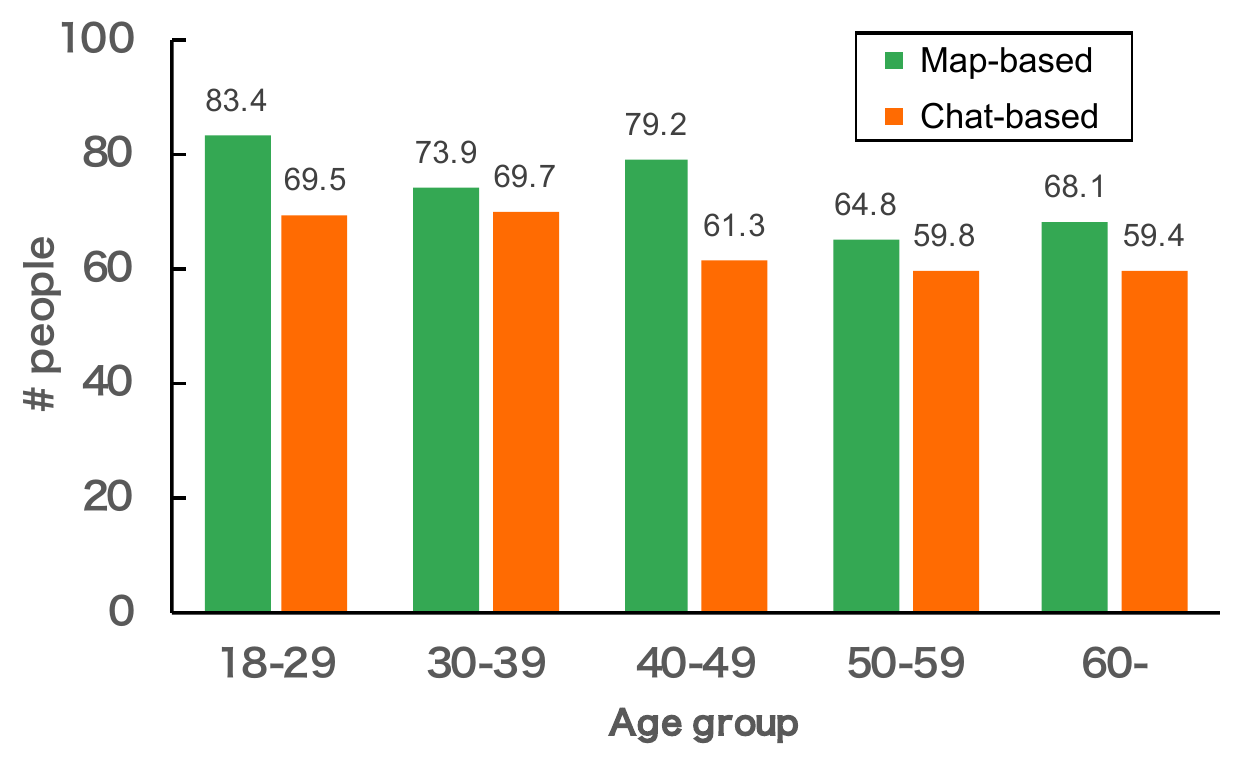}
            \hspace{1.6cm} {\small(a) Average SUS score by age group}
        \end{minipage}
        \begin{minipage}{0.38\columnwidth}
            \centering
            \includegraphics[bb=0 0 229 219, width=\columnwidth]{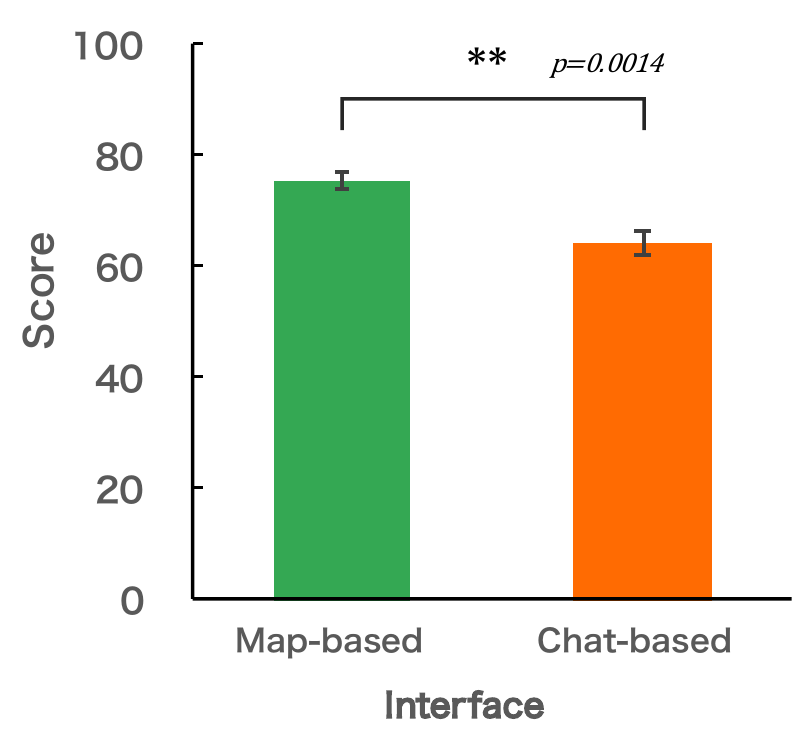}
            \hspace{1.6cm} {\small(b) Total average score}
        \end{minipage}
    \end{tabular}
    \caption{Summary of the application usability evaluation using SUS score}
    \label{fig:lgex-sus}
\end{figure}

\begin{figure}[t]
    \centering
    \includegraphics[bb= 0 0 773 292, width=\columnwidth]{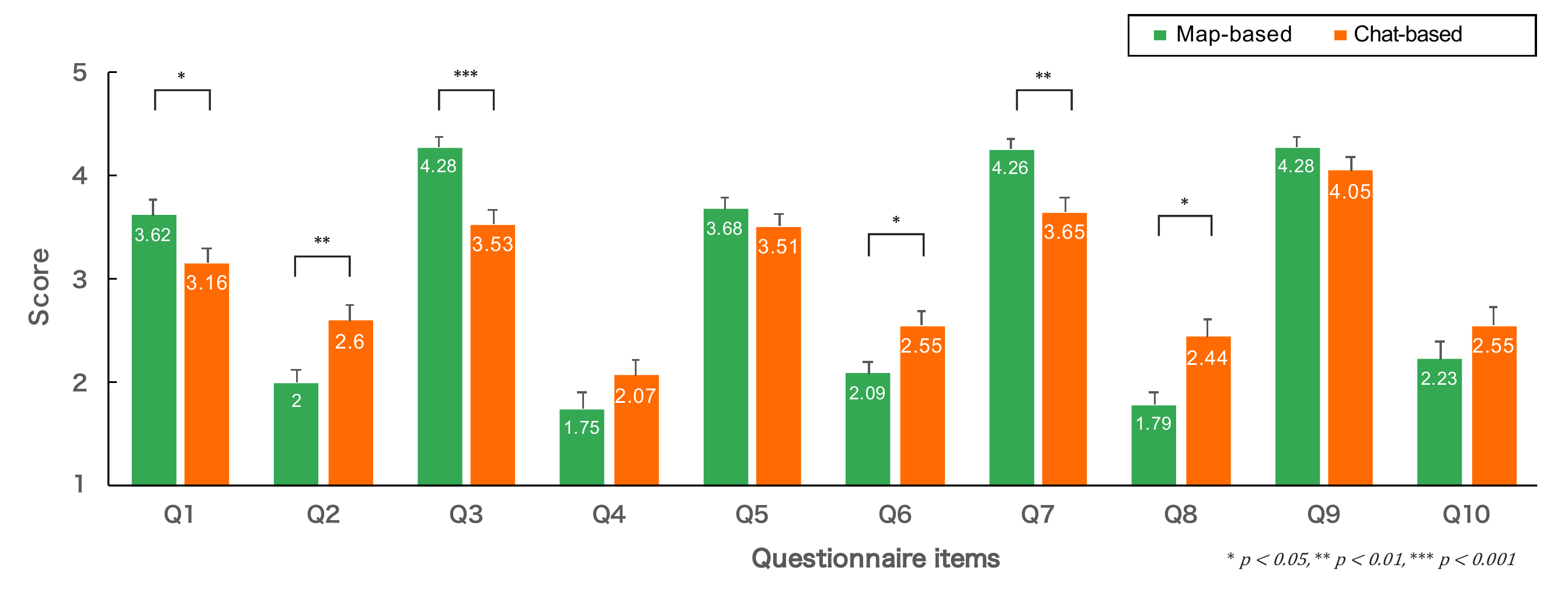}
    \caption{The average score for each of the ten items in each interface}
    \label{fig:lgex-lgex-sus-items}
\end{figure}

\noindent
\textbf{Application usability:}
The summary of the applixation usability evaluation using SUS score is shown in \figref{lgex-sus}; (a) shows the average SUS score by age group and (b) shows the overall average score of the answers.
The average and median SUS scores for the map-based interface were 75.6 and 75.0 ($S.D. = 12.7$), and for the chat-based interface were 64.3 and 67.5 ($S.D. = 17.6$).
In order to clarify whether there is a significant difference between them, the Mann-Whitney U test was performed since each SUS score did not follow normality in each interface and a significant difference was found between them ($p < 0.01$).
That is, the map-based interface was found to be a significantly more usable interface than the chat-based interface.
Next, in order to clarify the items that affected the difference in usability, the Man-Wittny U test was conducted to the responses  for each of  ten items. 
\figref{lgex-lgex-sus-items} shows the average score for each of the ten items in each interfaces, and items for which significant differences were found are indicated by asterisks on the bars. 
As a result, significant differences were found for questionnaire items Q1 ($p < 0.05$), Q2 ($p < 0.01$), Q3 ($p < 0.001$), Q6 ($p < 0.05$), Q7 ($p < 0.01$), and Q8 ($p < 0.05$).
Items Q2, Q3, and Q8 are related to complexity of application, and the scores for all items were more positive for the map-based interface. Item Q6 is related to consistency of the application and the Q7 is related the need for training until they can use the app, and the all scores were more positive for the map-based interface as well.
Due to these results, the map-based interface scored higher in Q1 about if they want to use this app frequently.
On the other hand, the items that did not find significant differences were Q4 and Q10 regarding the needs for support, Q5 regarding the consistency of the app, and Q9 regarding the confidence for using the app.
That is, there was no difference in the degree to which the users could use the application confidently without support once they start using the application, regardless of the interface.

\noindent
\textbf{Impressions through the experiment}
The following is examples of the free description impressions through the experiment.
First we sum up the coments given from the participants who used map-based interface.
Participant P23(Female, 46) answered, \textit{``Thanks to the app, I was able to visit places for the first time and know the places where I want to go in the next time, and I enjoyed sightseeing.''}.
P32(Female, 32) responded, \textit{``I could know there are various tourist spots, but I sometimes could not concentrate on one tourist spot because I thought like ``I want to go here! and there too!''. However, I was able to continue to enjoy sightseeing in Nara without getting bored.''}
P70 (Male, 69) mentioned, \textit{``I've been to Nara before, but I was able to find out places I didn't know through this experiment. I felt that if I could enjoy sightseeing with this app in the future, I would want to go to more places.''}.
Some of participants gave us the another aspect of views, like \textit{``Group travel is often avoided with this Corona situation, but I think this app can be a new way to share the joy of sightseeing.''} from participant P69 (Female, 42).

The following is the feedback that we got from the chat-based interface.
P38 (Female, 22) responded, \textit{``I became attached to Nara, through this experiment. Additionally, I've been Nara several times before, but I became more familiar in this time. I felt a little lonely because I was sightseeing alone, but it was good to be able to go around at my own pace. In addition to famous sightseeing spots, I was able to visit tourist spots that I didn't know or passed by if they weren't displayed in the app, and it was good to study history.''}
P50 (Male, 57) answered, \textit{``I had a meaningful experience in a place I didn't know well.''}
P77 (Female, 46) mentioned, \textit{``I was able to meet new places and beautiful scenery, and after reading the explanation of the points, I became more and more interested in Nara, thanks to this experiment. I arrived at my destination with peace of mind even on narrow roads, thanks to this app. I would like you to make it at other tourist spots. I think it would be great if a multilingual version was made and could be used by foreign tourists.''}
In addition, as a characteristic opinion in the chat-based interface, the following answer is obtained from Participant P49 (Female, 22) ; \textit{``Even though I was sightseeing alone, it was fun to feel like I wasn't alone while using the application.''}.

As mentioned above, most of the impressions obtained through the experiment were positive, but some participants gave us the following opinions.
Participant P95 (Male, 34, Map-based) mentioned, \textit{``With the points and rankings displayed on the app, I tried my best and walked too much.''}.
Participant P61 (Female, 20, Chat-based) described, \textit{``It was better for me to have no guidance. However, I think it was good that game elements such as ranking format were incorporated.''}

\subsection{Correlation between Behaviours and User Types}
Here, we clarify whether there are differences in data collection characteristics, tourism satisfaction, and interface preferences depending on the personality and user type of the participants.
The responses obtained by the 5-likert scale in the post-survey is used as an interval scale, and Pearson's product-moment correlation and test of Non-correlations will be used. 
The significance level is set at $ p<0.05 $, and $ 0.1>p>0.05 $ is considered as marginally significant.
First, we discuss the correlation with the data collection tendency.
We found a weak negative correlation and a marginally significant with Free Spirit in the map-based interface ($r=-0.26, p=0.06$).
In the total number of postings including free postings, there was also a weak negative correlation and significant difference with Free Spirit ($r=-0.38, p<0.01$).
When calculated the correlation with all free postings, no correlation or significant difference was found. 
However, a weak positive correlation and significant difference between the number of free postings and Philanthropist was found when the test was performed on the number of free postings of participants who had posted at least once ($r=0.34, p<0.01$). This tendency was similar when tested with the map-based ($r=0.40, p<0.01$) and chat-based ($r=0.32, p=0.05$) interfaces respectively.

Next, we describe the correlation with tourism satisfaction.
In the chat-based interface, a weak negative correlation and a marginally significant were found between tourism priority and Free Spirit ($r=-0.24, p=0.08$).
In terms of tourism enjoyment, there was a weak positive correlation and a significant difference in the attribute value of Player in the Map-based interface ($r=0.37, p<0.01$).

Finally, we discuss the correlation with interface preference.
Weak positive correlation and marginally significant between interface preferences and Player were found for the map-based interface ($r=0.25, p=0.07$).
In the chat-based interface, a weak positive correlation and a marginally significant were found with Achiever ($r=0.26, p=0.05$).

\section{Discussion}
In this section, we discuss the answers to our research questions based on the results obtained through the large scale experiment.

\noindent
\textbf{RQ1: How does the different task allocation interfaces affect the quantity and quality of dynamic tourism information collection?}

Regarding the quantity of the collected data, the map-based interface could collect about 1.4 times, and we found a significant difference between the interfaces. 
In the map-based interface, all the spots are visible at the same time, so it is possible to make a detour to other mission on the way to a certain destination. On the other hand, only the spots that have been decided in the chat dialogue are shown on the map in the chat-based interface.
The difference in the amount of data is attributed to the difference in the characteristics of the interface.
Another possible reason is that the chat-based interface requires more procedures than the map-based interface, as shown by the significant difference in the complexity of application in the SUS evaluation.
On the other hand, the chat-based interface could collect the high-demand data preferentially and efficiently. 
This result is caused by the fact that the chat-based interface uses an algorithm that prioritizes requests for spots with high points (high information demand) based on location information.

\begin{figure}[b]
    \centering
        \begin{tabular}{c}
        \begin{minipage}{0.45\columnwidth}
            \centering
            \includegraphics[bb=0 0 422 422, width=\columnwidth]{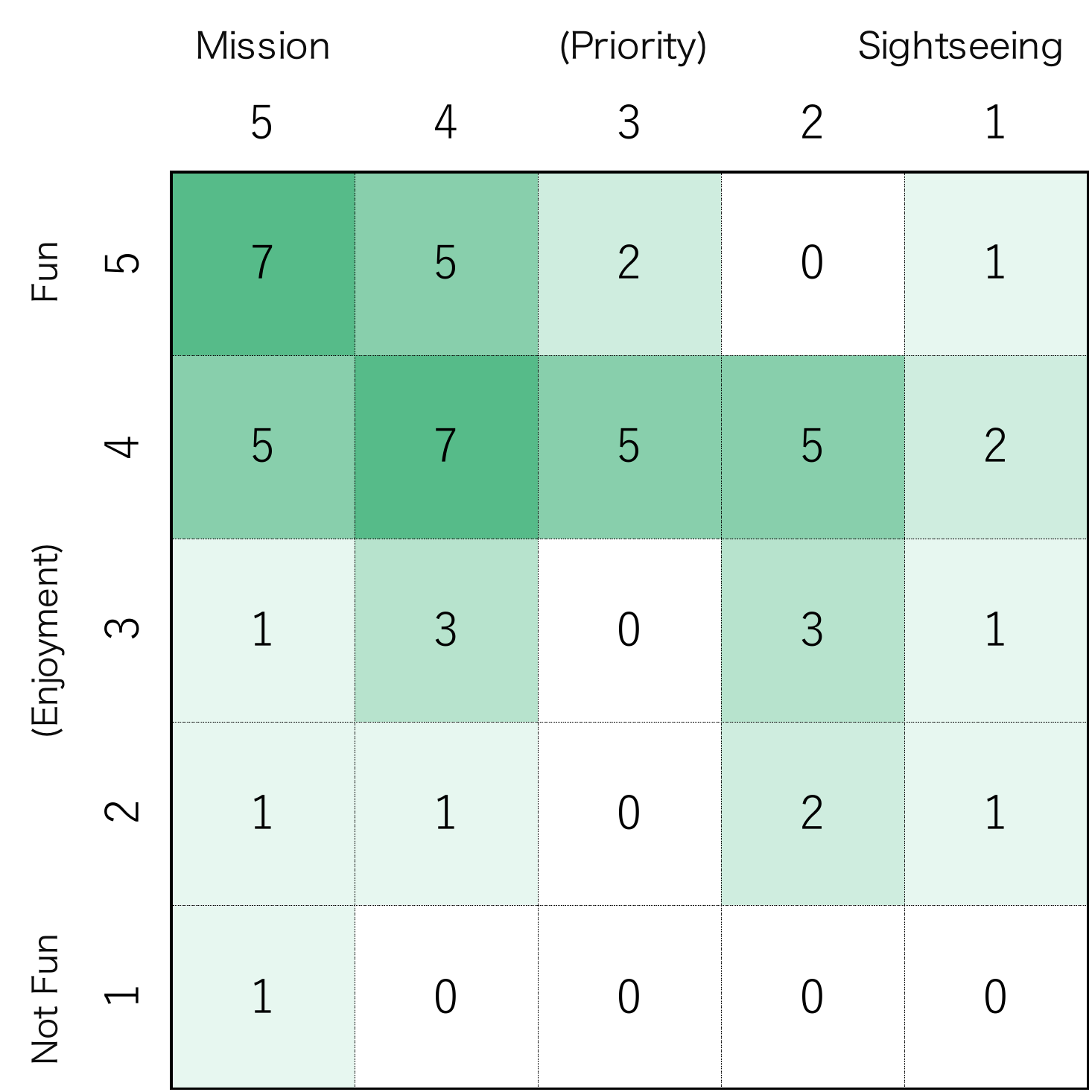}
            \hspace{1.6cm} {\small(a) Map-based}
        \end{minipage}
        \hspace{0.5cm}
        \begin{minipage}{0.45\columnwidth}
            \centering
            \includegraphics[bb=0 0 422 422, width=\columnwidth]{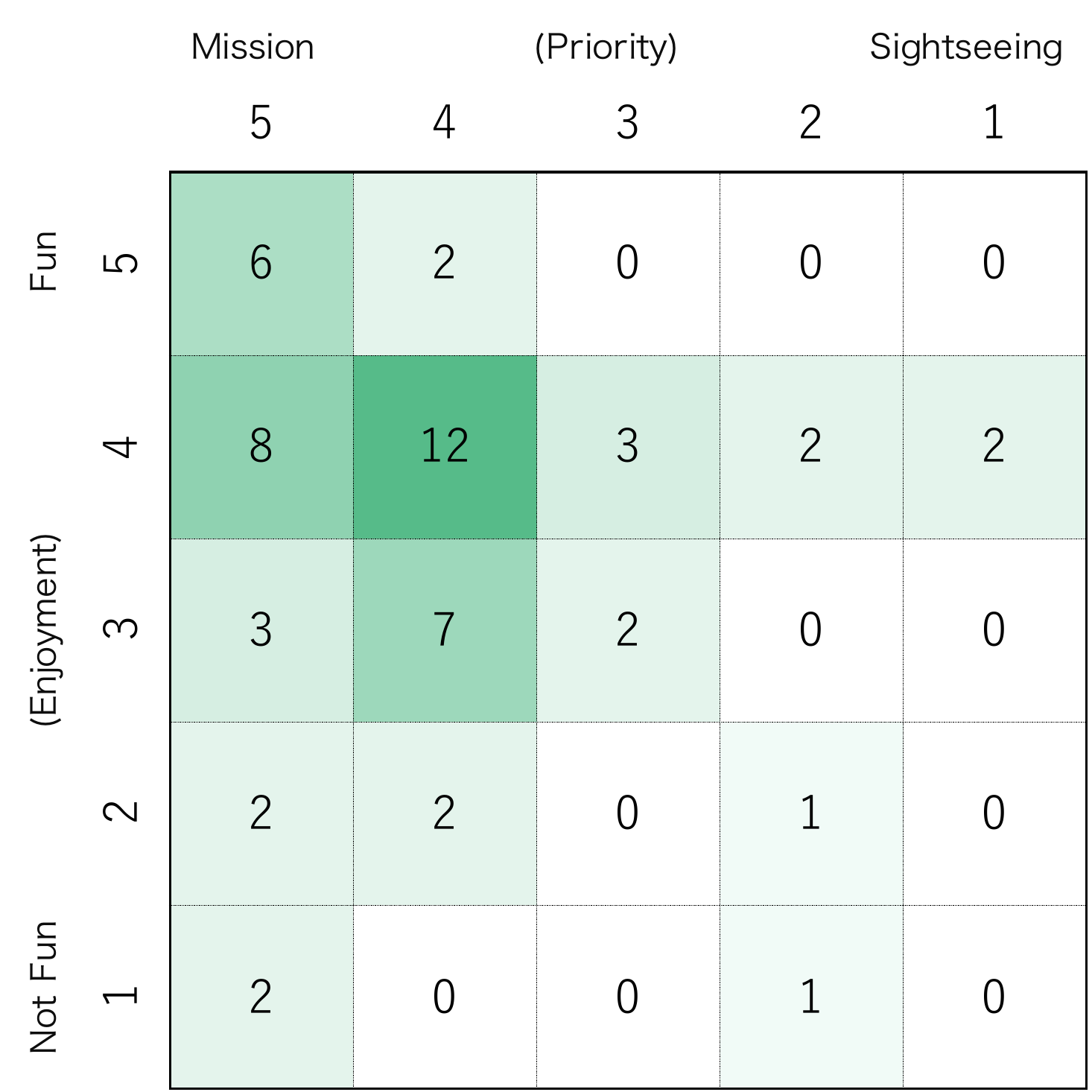}
            \hspace{1.6cm} {\small(b) Chat-based}
        \end{minipage}
    \end{tabular}
    \caption{The correspondence table between Q1:priority and Q2:enjoyment}
    \label{fig:lgex-corr-tab}
\end{figure}

\noindent
\textbf{RQ2: Do the different task allocation interfaces have an impact on tourism satisfaction of the tourists?}

From questionnaire Q1 and Q2, it was found that the mission was significantly more prioritized in the chat-based interface, but the interface difference did not affect the impact on the enjoyment of sightseeing.
The correspondence table between Q1: priority and Q2: enjoyment is shown in \figref{lgex-corr-tab}.
This result shows that the map-based interface has become a factor that makes sightseeing more enjoyable while balancing sightseeing and missions.
The same tendency can be seen in the chat-based interface, but the participants more prioritized to the mission. 
This result suggests that a certain sense of responsibility might be generated through agent interaction in chat.

\noindent
\textbf{RQ3: Is there a relationship between tourism information collection efficiency and interface preference, and gamification user type?} 

First, we describe the difference in data collection efficiency by user type.
A negative correlation was found between the number of mission posts and the Free Spirit attribute value in the map-based interface.
Free Spirit is a user type that is motivated by autonomy and is not constrained by external control.
It is assumed that participants with high autonomy might grasp missions as external controls and tend to perform them less frequently, while participants with relatively low autonomy tend to follow the missions and perform them more frequently.
Additionally, there was a negative correlation between Free Spirit and priority of sightseeing although in chat-based interface.
From these results, it is considered that the tendency to prioritize tourism and missions differs depending on the degree of autonomy, and participants with higher autonomy are more likely to prioritize their own tourism, resulting in a decrease in the number of posts.
On the other hand, there was a positive correlation between Philanthropist attribute value and the number of free posting, which allows participants to actively post at their own timing during sightseeing and obtains fewer points.
Philanthropists focus on purpose as their motivation and tend to act altruistically without the extrinsic rewards.
The number of free postings for the purpose of sharing the situation in the tourist attractions with the other participants is quite large, as shown in the open-ended responses for the purpose of free posting, such as ``when I find a place that is dangerous for people using the app or something I have never seen before'' (P2, Female, 38, Map-based), ``when I find a place that I want everyone to visit. (P25, Male, 28, Chat-based).
These factors suggest that users with high Philanthropists' attribute values tend to post more in order to share their situations at sightseeing spots with other participants by free posting and solving the timeline.
As examples of the reason for free posting, we obtained such comments, ``To tell people about dangerous places. Or when I find something I have never seen before.'' (P2, Female, 38, Map-based), `` When I found a place that I wanted everyone to visit.
 '' (P25, Male, 28, Chat-based).
These results suggest that participants with high Philanthropist attribute value tend to post more in order to share their situations at tourist attractions to other participants through free posts and timelines.

Next, we discuss the relationship between interface preference and user type.
There was a positive correlation with the Player attribute value in the map-based interface.
Players are mainly motivated by extrinsic rewards and will try to earn rewards from the system regardless of the type of activity.
In the map-based interface, all missions can be seen on a map, and they can see at a glance that the points to be obtained differ depending on the spot. Therefore, it is expected to stimulate the motivation of Player users who try to obtain higher points. 
We found also a positive correlation between the enjoyment of sightseeing and the Player attribute value in map-based interface.
That is, participants with a high Player value visit spots where they can get higher points, and enjoy sightseeing more while getting points, which may have increased their preference for the map-based interface.
On the other hand, there was a positive correlation with Achiever attribute value in the chat-based interface.
Achiever is mainly motivated by competence and seeks to perform the task given by the system.
In the chat-based interface, the agent character asks the participant to go to the spot where the system needs information as needed.
Therefore, it is considered that participants with high Achiever attribute value are more motivated to complete the given missions one after another, which in turn increases their preference for the chat-based interface.

\noindent
\textbf{RQ4: What is the impact of different communication style sentences in a chat-based interface?}

47~\% of the participants noticed a difference in the agent's communication style, showing that a large number of participants were aware of these subtle changes. Moreover, the results show that there is a clear preference in the \textit{elaborateness} dimension, both when looking at all participants and when looking at only those participants who noticed a difference in the agent's interaction. The participants significantly preferred the elaborate communication style over the concise one. This shows that, in general, more detailed information is preferred for the task at hand. However, this does not apply to all participants. Some of them clearly stated that they find a short and simple text easier to read on the smartphone while walking. Moreover, there is no preference for the \textit{indirectness} dimension. This is in line with the results presented in~\cite{bib:Miehle_usersatisfaction_2018}, that there is no general preference in the system's communication style and therefore the preference appears to be individual for every person.

\section{Conclusion}
The purpose of this paper was to investigate the effects of task allocation interfaces and user types on tourist information collection efficiency, tourist behavior, and tourist satisfaction in a gamified participatory sensing for tourists.
We designed and implemented two types of task allocation interfaces (map-based and chat-based), and we used four different communication styles based on two dimensions, \textit{elaborateness} and \textit{indirectness}, to elucidate the appropriate dialogue requests in the chat-based interface.
For gamification user modeling, we introduced the Hexad gamification user type which defined by Tondello et al.~\cite{bib:hexed-usertype_tondello_2016}.
Then, we set four research questions and these were clarified through a large-scale sightseeing experiments with 108 ordinary people aged between 19 and 71 in Nara.
We found the following through from the experiment.
The absolute number of contributions was about 1.4 times greater for the map-based interface than the chat-based interface, but it was more efficient in obtaining the data required by the system.
In addition, there was no significant difference in the tourism satisfaction between the two interfaces. However, we found different trends for the contribution to sensing and the interface preference by user type.
As for the free posting, there was a positive and weak correlation between the number of contributions and the Philanthropist attribute value for the participants who posted more than one contribution. That is, the higher the value of Philanthropy, the higher the number of free contributions.
There was no significant difference in the impact of the different interfaces on the enjoyment of tourism. 
On the other hand, the usability of the application was higher for the map-based interface in terms of SUS score.
When evaluating the different communication styles of the chat-based interface, we could show that 47~\% of the participants noticed a difference in the agent's communication style. Hence, a large number of participants were aware of these subtle changes. Furthermore, the participants significantly preferred the elaborate communication style over the concise one. However, there is no preference for the \textit{indirectness} dimension. This shows that there is no general preference in the system's communication style and therefore the preference appears to be individual for every person. 
The overall impression of the experiment was generally positive, for example, that using minor tourist attractions as checkpoints would give a sense of serendipity to tourism. 

\bibliographystyle{acm}
\bibliography{references}  

\begin{thebibliography}{10}

\bibitem{bib:arakawa_gamification_2016}
{\sc Arakawa, Y., and Matsuda, Y.}
\newblock Gamification mechanism for enhancing a participatory urban sensing:
  survey and practical results.
\newblock {\em Journal of Information Processing 24}, 1 (2016), 31--38.

\bibitem{bib:sus}
{\sc Bangor, A., Kortum, P.~T., and Miller, J.~T.}
\newblock An empirical evaluation of the system usability scale.
\newblock {\em International Journal of Human^^e2^^80^^93Computer Interaction
  24}, 6 (2008), 574--594, https://doi.org/10.1080/10447310802205776.

\bibitem{bib:burke_participatory_2006}
{\sc Burke, J.~A., Estrin, D., Hansen, M., Parker, A., Ramanathan, N., Reddy,
  S., and Srivastava, M.~B.}
\newblock Participatory sensing.
\newblock {\em Center for Embedded Network Sensing\/} (2006).

\bibitem{bib:teenage_vanessa_2020}
{\sc Ces\'{a}rio, V., Petrelli, D., and Nisi, V.}
\newblock Teenage visitor experience: Classification of behavioral dynamics in
  museums.
\newblock In {\em Proceedings of the 2020 CHI Conference on Human Factors in
  Computing Systems\/} (New York, NY, USA, 2020), CHI '20, Association for
  Computing Machinery, p.~1^^e2^^80^^9313.

\bibitem{bib:deterding_mindtrek_2011}
{\sc Deterding, S., Dixon, D., Khaled, R., and Nacke, L.}
\newblock From game design elements to gamefulness: Defining "gamification".
\newblock In {\em Proceedings of the 15th International Academic MindTrek
  Conference: Envisioning Future Media Environments\/} (New York, NY, USA,
  2011), MindTrek '11, ACM, pp.~9--15.

\bibitem{bib:dbj}
{\sc {Development Bank of Japan Inc.}}
\newblock {Report of the Regional Planning Department: 2014 Survey of Travelers
  to Japan from Eight Asian Regions}.
\newblock \url{https://www.dbj.jp/en/pdf/investigate/etc/pdf/book1412_01.pdf},
  2014.

\bibitem{bib:Elhamshary_crowdmeter_percom_2018}
{\sc {Elhamshary}, M., {Youssef}, M., {Uchiyama}, A., {Yamaguchi}, H., and
  {Higashino}, T.}
\newblock Crowdmeter: Congestion level estimation in railway stations using
  smartphones.
\newblock In {\em 2018 IEEE International Conference on Pervasive Computing and
  Communications (PerCom)\/} (March 2018), pp.~1--12.

\bibitem{bib:Gretzel_SmartTourism_EM_2015}
{\sc Gretzel, U., Sigala, M., Xiang, Z., and Koo, C.}
\newblock Smart tourism: foundations and developments.
\newblock {\em Electronic Markets 25}, 3 (Sep 2015), 179--188.

\bibitem{bib:himari_homo_2013}
{\sc Hamari, J.}
\newblock Transforming homo economicus into homo ludens: A field experiment on
  gamification in a utilitarian peer-to-peer trading service.
\newblock {\em Electronic Commerce Research and Applications 12}, 4 (2013), 236
  -- 245.
\newblock Social Commerce- Part 2.

\bibitem{bib:hamari_gamification_hicss_2014}
{\sc {Hamari}, J., {Koivisto}, J., and {Sarsa}, H.}
\newblock Does gamification work? -- a literature review of empirical studies
  on gamification.
\newblock In {\em 2014 47th Hawaii International Conference on System
  Sciences\/} (Jan 2014), pp.~3025--3034.

\bibitem{bib:hamari_playertype_2014}
{\sc Hamari, J., and Tuunanen, J.}
\newblock Player types: A meta-synthesis.
\newblock {\em Transactions of the Digital Games Research Association 1\/} (03
  2014), 29--53.

\bibitem{bib:hidaka_SmartCities_2020}
{\sc Hidaka, M., Kanaya, Y., Kawanaka, S., Matsuda, Y., Nakamura, Y., Suwa, H.,
  Fujimoto, M., Arakawa, Y., and Yasumoto, K.}
\newblock On-site trip planning support system based on dynamic information on
  tourism spots.
\newblock {\em Smart Cities 3}, 2 (2020), 212--231.

\bibitem{bib:iso-tour_smartcities_2020}
{\sc Isoda, S., Hidaka, M., Matsuda, Y., Suwa, H., and Yasumoto, K.}
\newblock Timeliness-aware on-site planning method for tour navigation.
\newblock {\em Smart Cities 3}, 4 (2020), 1383--1404.

\bibitem{bib:jaimes_survey_ieee_2015}
{\sc {Jaimes}, L.~G., {Vergara-Laurens}, I.~J., and {Raij}, A.}
\newblock A survey of incentive techniques for mobile crowd sensing.
\newblock {\em IEEE Internet of Things Journal 2}, 5 (Oct 2015), 370--380.

\bibitem{bib:personality-gamification_jia_2016}
{\sc Jia, Y., Xu, B., Karanam, Y., and Voida, S.}
\newblock Personality-targeted gamification: A survey study on personality
  traits and motivational affordances.
\newblock In {\em Proceedings of the 2016 CHI Conference on Human Factors in
  Computing Systems\/} (New York, NY, USA, 2016), CHI '16, Association for
  Computing Machinery, p.~2001^^e2^^80^^932013.

\bibitem{bib:gamified_kawanaka_mdpi_2020}
{\sc Kawanaka, S., Matsuda, Y., Suwa, H., Fujimoto, M., Arakawa, Y., and
  Yasumoto, K.}
\newblock Gamified participatory sensing in tourism: An experimental study of
  the effects on tourist behavior and satisfaction.
\newblock {\em Smart Cities 3}, 3 (2020), 736--757.

\bibitem{bib:kim_chatbot_2019}
{\sc Kim, S., Lee, J., and Gweon, G.}
\newblock Comparing data from chatbot and web surveys: Effects of platform and
  conversational style on survey response quality.
\newblock In {\em Proceedings of the 2019 CHI Conference on Human Factors in
  Computing Systems\/} (New York, NY, USA, 2019), CHI '19, Association for
  Computing Machinery, p.~1^^e2^^80^^9312.

\bibitem{bib:lee_psy_behav_mdpi_2019}
{\sc Lee, B.~C.}
\newblock The effect of gamification on psychological and behavioral outcomes:
  Implications for cruise tourism destinations.
\newblock {\em Sustainability 11}, 11 (2019).

\bibitem{bib:malone_motivating_1981}
{\sc Malone, T.~W.}
\newblock Toward a theory of intrinsically motivating instruction.
\newblock {\em Cognitive Science 5}, 4 (1981), 333--369.

\bibitem{bib:ninja_monkeys_2015}
{\sc Marczewski, A.}
\newblock {\em Even Ninja Monkeys Like to Play: Gamification, Game Thinking and
  Motivational Design}.
\newblock CreateSpace Independent Publishing Platform, 2015.

\bibitem{bib:yukimat_parmosense_arXiv_2021}
{\sc Matsuda, Y., Kawanaka, S., Suwa, H., Arakawa, Y., and Yasumoto, K.}
\newblock Parmosense: A scenario-based participatory mobile urban sensing
  platform with user motivation engine, 2021, 2102.05586.

\bibitem{bib:Miehle_estimation_2020}
{\sc Miehle, J., Feustel, I., Hornauer, J., Minker, W., and Ultes, S.}
\newblock Estimating user communication styles for spoken dialogue systems.
\newblock In {\em Proceedings of the 12th Language Resources and Evaluation
  Conference\/} (Marseille, France, May 2020), European Language Resources
  Association, pp.~540--548.

\bibitem{bib:Miehle_usersatisfaction_2018}
{\sc Miehle, J., Minker, W., and Ultes, S.}
\newblock Exploring the impact of elaborateness and indirectness on user
  satisfaction in a spoken dialogue system.
\newblock In {\em Adjunct Publication of the 26th Conference on User Modeling,
  Adaptation and Personalization\/} (New York, NY, USA, 2018), UMAP '18,
  Association for Computing Machinery, p.~165^^e2^^80^^93172.

\bibitem{bib:Benedikt_IJHCS_2017}
{\sc Morschheuser, B., Hamari, J., Koivisto, J., and Maedche, A.}
\newblock Gamified crowdsourcing: Conceptualization, literature review, and
  future agenda.
\newblock {\em International Journal of Human-Computer Studies 106\/} (2017),
  26 -- 43.

\bibitem{bib:narimoto_wayfinding_2018}
{\sc {Narimoto}, R., {Kajita}, S., {Yamaguchi}, H., and {Higashino}, T.}
\newblock Wayfinding behavior detection by smartphone.
\newblock In {\em 2018 IEEE 32nd International Conference on Advanced
  Information Networking and Applications (AINA)\/} (May 2018), pp.~488--495.

\bibitem{bib:Pragst_comstyle_2017}
{\sc Pragst, L., Minker, W., and Ultes, S.}
\newblock Exploring the applicability of elaborateness and indirectness in
  dialogue management.
\newblock In {\em IWSDS\/} (2017).

\bibitem{bib:deci_sdt_2000}
{\sc Ryan, R.~M., and Deci, E.~L.}
\newblock Self-determination theory and the facilitation of intrinsic
  motivation, social development, and well-being.
\newblock {\em American psychologist 55}, 1 (2000), 68.

\bibitem{bib:seaborn_gam_survey_elsevior_2015}
{\sc Seaborn, K., and Fels, D.~I.}
\newblock Gamification in theory and action: A survey.
\newblock {\em International Journal of Human-Computer Studies 74\/} (2015), 14
  -- 31.

\bibitem{bib:Sigala_Springer_2015}
{\sc Sigala, M.}
\newblock {\em Gamification for Crowdsourcing Marketing Practices: Applications
  and Benefits in Tourism}.
\newblock Springer International Publishing, Cham, 2015, pp.~129--145.

\bibitem{bib:Sigala_Ashgate_2012}
{\sc Sigala, M., Christou, E., and Gretzel, U.}
\newblock {\em Social Media in Travel, Tourism and Hospitality; Theory,
  Practice and Cases}.
\newblock Ashgate Publishing, Ltd., 01 2012.

\bibitem{bib:hexed-usertype_tondello_2019}
{\sc Tondello, G.~F., Mora, A., Marczewski, A., and Nacke, L.~E.}
\newblock Empirical validation of the gamification user types hexad scale in
  english and spanish.
\newblock {\em International Journal of Human-Computer Studies 127\/} (2019),
  95 -- 111.

\bibitem{bib:hexed-usertype_tondello_2016}
{\sc Tondello, G.~F., Wehbe, R.~R., Diamond, L., Busch, M., Marczewski, A., and
  Nacke, L.~E.}
\newblock The gamification user types hexad scale.
\newblock In {\em Proceedings of the 2016 Annual Symposium on Computer-Human
  Interaction in Play\/} (New York, NY, USA, 2016), CHI PLAY '16, Association
  for Computing Machinery, p.~229^^e2^^80^^93243.

\bibitem{bib:ueyama_gamification_2014}
{\sc Ueyama, Y., Tamai, M., Arakawa, Y., and Yasumoto, K.}
\newblock Gamification-based incentive mechanism for participatory sensing.
\newblock In {\em 2014 IEEE International Conference on Pervasive Computing and
  Communication Workshops (PerCom Workshops)\/} (2014), pp.~98--103.

\bibitem{bib:Xu_TM_2017}
{\sc Xu, F., Buhalis, D., and Weber, J.}
\newblock Serious games and the gamification of tourism.
\newblock {\em Tourism Management 60\/} (2017), 244 -- 256.

\end{thebibliography}

\newpage
\appendix
\section{Gamification User Types Hexad scale}

Hexad is a gamification user type model consisting of six types to personalize the gamification design according to the user's personality~\cite{bib:ninja_monkeys_2015, bib:hexed-usertype_tondello_2016}.
The Gamification User Types Hexad scale is a 24-items survey response scale to
score users’ preferences towards the six different motivations in the Hexad framework~\cite{bib:hexed-usertype_tondello_2019}.
We asked the participants to rate how well each item describes them in a 7-point Likert scale ; 1: Strongly disagree and 7: Strongly agree; without identifying the corresponding type. 
The user type scores are calculated that separately add the scores of the items corresponding to each subscale.
Basically, the one with the highest value in each subscale is used as the representative user type, but it is an archetypical categorization in Hexad framework.

\begin{table}[ht]
    \centering
    \caption{The Gamification User Types Hexad scale}
    \label{tab:questionnaire_hexad}
    \begin{tabular}{lcwl{0.69\columnwidth}}
        \toprule
        User Types	&	\#	&	English Items	\\
        \midrule
        Philanthropist	&	P1	&	It makes me happy if I am able to help others.	\\
        	&	P2	&	I like helping others to orient themselves in new situations.\\
        	&	P3	&	I like sharing my knowledge.	\\
        	&	P4	&	The well being of others is important to me.	\\ \hline
        Socialiser	&	S1	&	Interacting with others is important to me.	\\
        	&	S2	&	I like being part of a team.	\\
        	&	S3	&	It is important to me to feel like I am part of a community.\\
        	&	S4	&	I enjoy group activities.	\\\hline
        Free Spirit	&	F1	&	It is important to me to follow my own path.	\\
        	&	F2	&	I often let my curiosity guide  me.	\\
        	&	F3	&	I like to try new things.	\\
        	&	F4	&	Being independent is important to me.	\\\hline
        Achiever	&	A1	&	I like defeating obstacles.	\\
        	&	A2	&	It is important to me to always carry out my tasks completely. \\
        	&	A3	&	It is difficult for me to let go of a problem before I have found a solution. \\
        	&	A4	&	I like mastering difficult tasks.	\\\hline
        Player	&	R1	&	I like competitions where a prize can be won.	\\
        	&	R2	&	Rewards are a great way to motivate me.	\\
        	&	R3	&	Return of investment is important to me	\\
        	&	R4	&	If the reward is sufficient I will put in the effort.	\\\hline
        Disruptor	&	D1	&	I like to provoke.	\\
        	&	D2	&	I like to question the status quo.	\\
        	&	D3	&	I see myself as a rebel.	\\
        	&	D4	&	I dislike following rules.	\\
        \bottomrule
    \end{tabular}
\end{table}

\section{System Usability Scale (SUS) \label{appx:sus}}
The SUS is a simple, ten-item scale giving a global view of subjective assessments of usability~\cite{bib:sus}.
It is calculated with following questionnaire items.
The questionnaire is answered with 5 point Likert scale; 1: Strongly disagree and 5: Strongly agree.

\begin{enumerate}
    \item I think that I would like to use this system frequently.
    \item I found the system unnecessarily complex.
    \item I thought the system was easy to use.
    \item I think that I would need the support of a technical person to be able to use this system.
    \item I found the various functions in this system were well integrated.
    \item I thought there was too much inconsistency in this system.
    \item I would imagine that most people would learn to use this system very quickly.
    \item I found the system very cumbersome to use.
    \item I felt very confident using the system.
    \item I needed to learn a lot of things before I could get going with this system.
\end{enumerate}

To calculate the SUS score, first sum the score contributions from each item. Each item's
score contribution will range from 0 to 4. For items 1,3,5,7,and 9 the score contribution is the
scale position minus 1. For items 2,4,6,8 and 10, the contribution is 5 minus the scale
position. Multiply the sum of the scores by 2.5 to obtain the overall value of SUS.






\end{document}